\documentclass{emulateapj}

\usepackage{epsfig}
\usepackage{amsmath}
\usepackage{natbib}

\newcommand{\degree}{\ensuremath{^\circ}}
\newcommand {\kmskpc}{\ensuremath{\text{km s}^{-1} \text{kpc}^{-1}}}
\begin{document}
\title{Uncovering the Origins of Spiral Structure By Measuring Radial Variation in Pattern Speeds}
\author{Sharon E. Meidt and Richard J. Rand}
\affiliation{Department of Physics and Astronomy, \\University of New Mexico, 800 Yale Blvd Northeast, Albuquerque, NM 87131}
\author{Michael R. Merrifield}
\affiliation{School of Physics $\&$ Astronomy, \\University of Nottingham, University Park, Nottingham, NG7 2RD, UK}

\date{\today}

\begin{abstract}

Current theories of spiral and bar structure predict a variety of pattern speed behaviors, calling for detailed, direct measurement of the radial variation of pattern speeds.  
 Our recently developed Radial Tremaine-Weinberg (TWR) method allows this goal to be achieved for the first time.
Here we present TWR spiral pattern speed estimates for M101, IC 342, NGC 3938 and NGC 3344 in order to investigate whether spiral structure is steady or winding, whether spirals are described by multiple pattern speeds, and the relation between bar and spiral speeds.  Where possible, we interpret our pattern speeds estimates according to the resonance radii associated with each (established with the disk angular rotation), and compare these to previous determinations.  
By analyzing the high-quality HI and CO data cubes available for these
galaxies, we show that it is possible to determine directly
multiple pattern speeds within these systems, and hence identify the
characteristic signatures of the processes that drive the spiral
structure.  Even this small sample of galaxies reveals a surprisingly
complex taxonomy, with the first direct evidence for the presence of
resonant coupling of multiple patterns found in some systems, and the measurement of a simple single
pattern speed in others.  
Overall, this study demonstrates that we are now in a position to uncover more of the
apparently complex physics that lies behind spiral structure.  
\end{abstract}

\keywords{galaxies: individual (M101, IC 342, NGC 3938, NGC3344) --
galaxies: spiral --
galaxies: kinematics and dynamics -- 
galaxies: structure --
methods: numerical}
\section{Introduction}
The angular rate at which bar and spiral structures rotate, the pattern speed, is a parameter appreciated for its capacity to reveal information about the appearance, properties and evolution of galaxy disks.  Since the pattern speed determines the rate at which gas and stars encounter these structures, it provides a description of their influence on star formation, the processing of the interstellar medium and the distribution of metals (e.g. \citealt{rand93}; \citealt{knapen}; \citealt{henry}), for example, as well as the redistribution of mass in the disk.  Bar torques depend on 
the pattern speed, and therefore gas inflow rates \citep{quillenTorque} do, as well.  
In early-type barred galaxies, pattern speeds have also been used to investigate the central Dark Matter (DM) content in these systems (e.g. \citealt{debs}; \citealt{corsini}; but see \citealt{dubinski}).  \\
\indent The pattern speed moreover supplies a fundamental characterization with which to investigate the origin and evolution of large-scale structures, one of the prime unresolved issues in the dynamics and evolution of galaxy disks.  
With knowledge of bar and spiral or the radial variation of spiral pattern speeds in a given galaxy it is possible to determine whether spiral structure is steady or winding, the domain and number of patterns that can be sustained in a disk, and the relation between bar and spiral pattern speeds. Direct pattern speed measurements, however, have been elusive.\\
\indent Although grand-design spirals have been observationally linked to the presence of bars or companions \citep{kn}, it remains unclear whether spirals are long-lived density waves persisting over many revolutions (i.e. \citealt{linshu}; \citealt{bertin}), or rapidly-evolving, transient features, as found in several simulations (\citealt{sc}; \citealt{thomasson}; \citealt{sk}).  \\
\indent It is also uncertain how waves can exist over large fractions of a disk radius, as observed.  Since the resonances (inner and outer Lindblad; ILR and OLR) between which waves are expected to propagate in general do not span a large range in radius, waves of different speeds and structure may occupy distinct radial zones \citep{masset2}.   
In barred galaxies, the apparent alignment between bar and spiral (as in M83, where the two-armed spiral emanates from the bar ends), initially taken as evidence for a common pattern speed, does not occur in general \citep{ss88}.  Furthermore, where the bar ends near its Corotation Resonance (CR) (e.g. as found in early-types; \citealt{corsini}), a spiral generated with the same speed would lie mostly outside the CR (i.e. with dust lanes along the convex side of the arms), which is also uncommon.\\
\indent In the theory of ``mode coupling'' (\citealt{syg}; \citealt{masset2}), multiple patterns in different radial zones are linked, such that the resulting wave structure can extend over a larger radial range than is possible for a single pattern. In this scenario, the CR of an inner pattern overlaps with the ILR of an outer pattern, and at this overlap energy and angular momentum are efficiently transferred outward in the disk.  This was first demonstrated between simulated bars and spirals by \citet{masset2} and later by \citet{rs}, who find, in addition, that other resonance overlaps, such as the coincidence of the spiral's inner 4:1 resonance with the bar CR, are also effective.  Given that the radial extent of resonant orbits can be quite large, these overlappings can be spatially broad, and frequency diagrams can therein be suitably revealing.  However, not all such resonance overlaps are accompanied by signs of true mode-coupling (i.e. boosted beat modes detectable in simulation power spectra at the overlap; \citet{masset2}). \\
\indent Spiral-spiral mode coupling may also occur, typified, perhaps, by the transition commonly observed between a strong two-armed spiral and more complex structure at some radius.  In their simulations, \citet{rs} find evidence for spiral structure in the absence of a bar, spiral-spiral mode coupling, and multiple pattern speeds without mode coupling.\\
\indent To observationally investigate these possibilities, we have undertaken a project to measure spiral pattern speeds and their radial variation using the Radial Tremaine-Weinberg method developed by \citet{mrm}, subsequently tested on simulations \citep{meidt08a}; hereafter Paper I), and recently applied to a real spiral in the grand-design galaxy M51 (\citealt{meidt08b}, hereafter Paper II).  
Like its traditional counterpart--the TW method for measuring a single, constant pattern speed \citep{tw84}--the TWR calculation yields measurements of pattern speeds using observationally accessible quantities based on a requirement of continuity.  The method is direct, and so overcomes several of the obstacles in reliably estimating pattern speeds with other, indirect methods (e.g. resonance identification or modeling; \citealt{elm89}, \citealt{elm96}, \citealt{raut2005}, \citealt{gb}).  \\
\indent  Moreover, the TWR method provides a resolution for those applications of the TW method where pattern speed estimates show a clear, systematic departure from a single value (e.g. \citealt{zrm04} and \citealt{mrm}).  Particularly in disks with multiple or extended structures, this behavior implies that the pattern speed is not constant either because it varies temporally or spatially. By allowing the pattern speed to vary in the radial direction, the TWR method explicitly allows for the possibility that a galaxy may contain a number of distinct features at different radii, such as bars and spiral arms, each with their own pattern speeds. It also makes possible the detection of spiral winding and hence the estimation of the lifetime of a galaxy's current pattern.\\
\indent Our aim with this paper is to expand to a larger sample of galaxies the TWR analysis recently applied to the grand-design spiral galaxy M51 (Paper II).  Here, as there, we apply the method according to the prescription outlined in Paper I.  The calculation employs regularization, which smooths otherwise intrinsically noisy solutions through the use of a prior models of the radial dependence of the pattern speed, and affords straightforward tests for bar-spiral and spiral-spiral relations, and spiral winding.  \\
\indent Also as in Paper II, for use as a kinematic tracer we consider observations of the ISM, which have become the standard choice of spiral tracer for meeting the continuity requirement of the method.  
The application of the TW and TWR methods to CO and HI observations of spiral galaxies (e.g. \citealt{westpfahl}; \citealt{rw04}; \citealt{mrm}) avoids significant problems with the stellar component in these systems, namely the faintness of the old stellar disk, and the effects of star formation and obscuration by dust in spiral arms by which the application of the continuity equation is invalidated.  \\
\indent Furthermore, where the ISM is dominated by either the molecular (traced by CO) or atomic gas phases, conversion among phases can be assumed to occur at low levels on orbital timescales such that, together with the low true efficiency of star formation in spirals, the dominant component arguably obeys continuity (e.g. \citealt{zrm04}; \citealt{rw04}).  In addition, as an improvement on previous ISM-based TW spiral studies, here we analyze both molecular gas observations from the BIMA SONG (\citealt{helfer}; maps include zero-spacing flux information) and archival 21-cm emission data tracing the atomic hydrogen phase.  (The flux information at the largest scales in all of the HI cubes considered here is comparable to that achieved with single-dish observations; see references in $\S\S$\ref{sec:samp5457}-\ref{sec:samp3344}). Our consideration of both CO and the more extended HI here serves two main purposes.  In galaxies that are not molecule-dominated over the extent of the detectable CO emission, the HI supplements H$_2$ to establish a total particle, continuity-obeying tracer.  The radial range of detectable pattern speeds is also increased in this way.\\
\indent Part of our treatment, in this case, entails an investigation (where possible) into the sensitivity of TWR solutions to the CO-to-H$_2$ conversion factor $X$ adopted in combining the data, which has been suggested to vary linearly with metallicity (e.g. \citealt{boselli}).  Following \citet{zrm04} and Paper II, we consider the effect of variation in $X$ with radius (but not the possibility, for example, of arm-interarm variations). 
We also take into consideration distortions or warps observable in the outer HI disks in our sample (described in $\S\S$\ref{sec:sample}-\ref{sec:samp3344}), which violate of one of the main TW assumptions (namely, that the disk is flat; \citealt{tw84}).  \\
\indent Based on the a priori models described in $\S$\ref{sec:results}, the best-fit pattern speed solutions calculated for each galaxy are presented in $\S\S$ \ref{sec:results5457}-\ref{sec:results3344}.  There, the resonances associated with each, which we identify through comparison with angular rotation curves, serve as a main informant of our results; our findings are motivated by, and compared with, the large body of work linking resonances to the dynamics and morphology of disks.   We also discuss the limitations and sensitivities of the TWR calculation (described in Paper I) as applied to each galaxy and summarize our results in $\S$4.
\section{The Sample}
\subsection{\label{sec:sample}Selection and Overview}
\begin{figure*}
\begin{center}
\epsscale{1.}
\plottwo{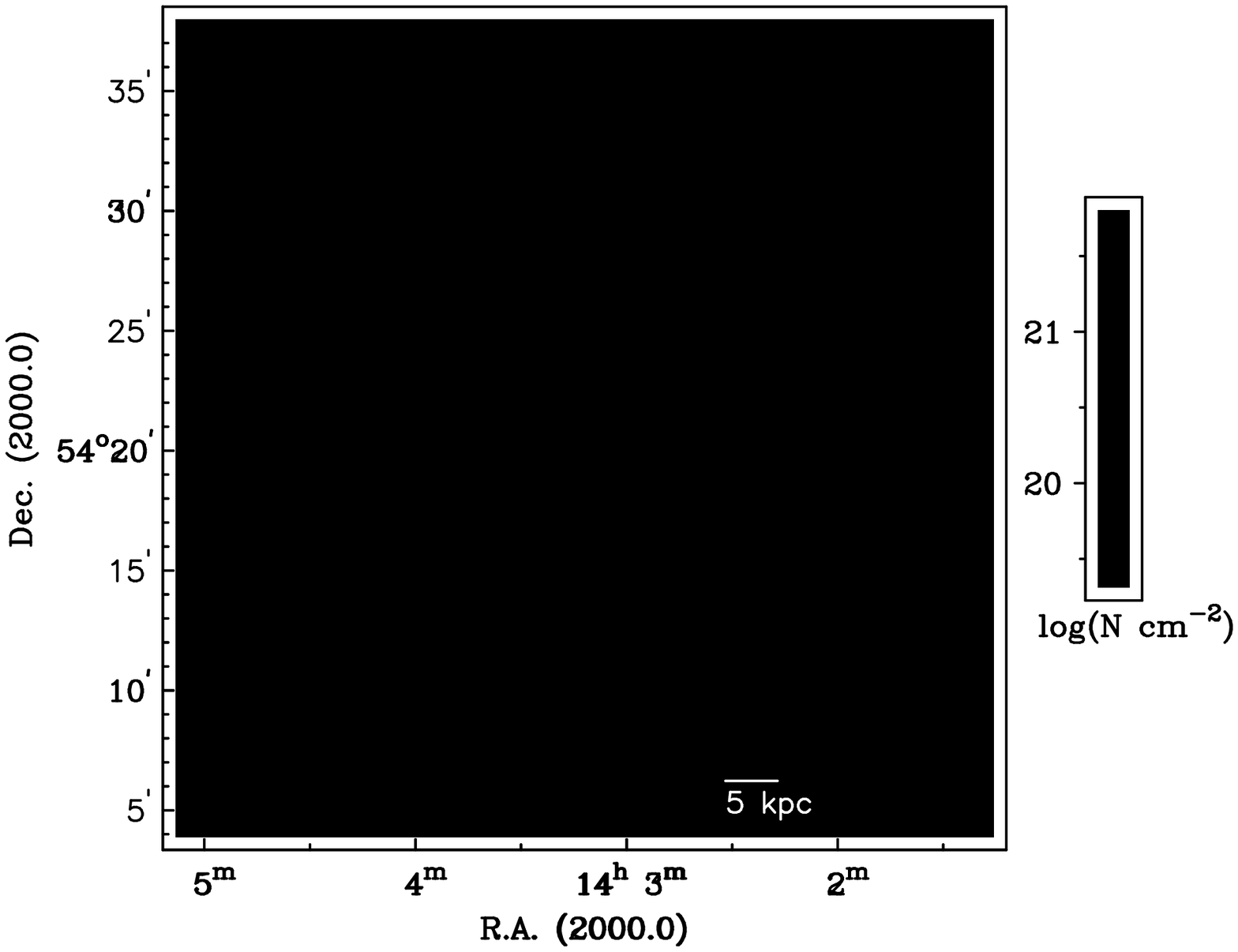}{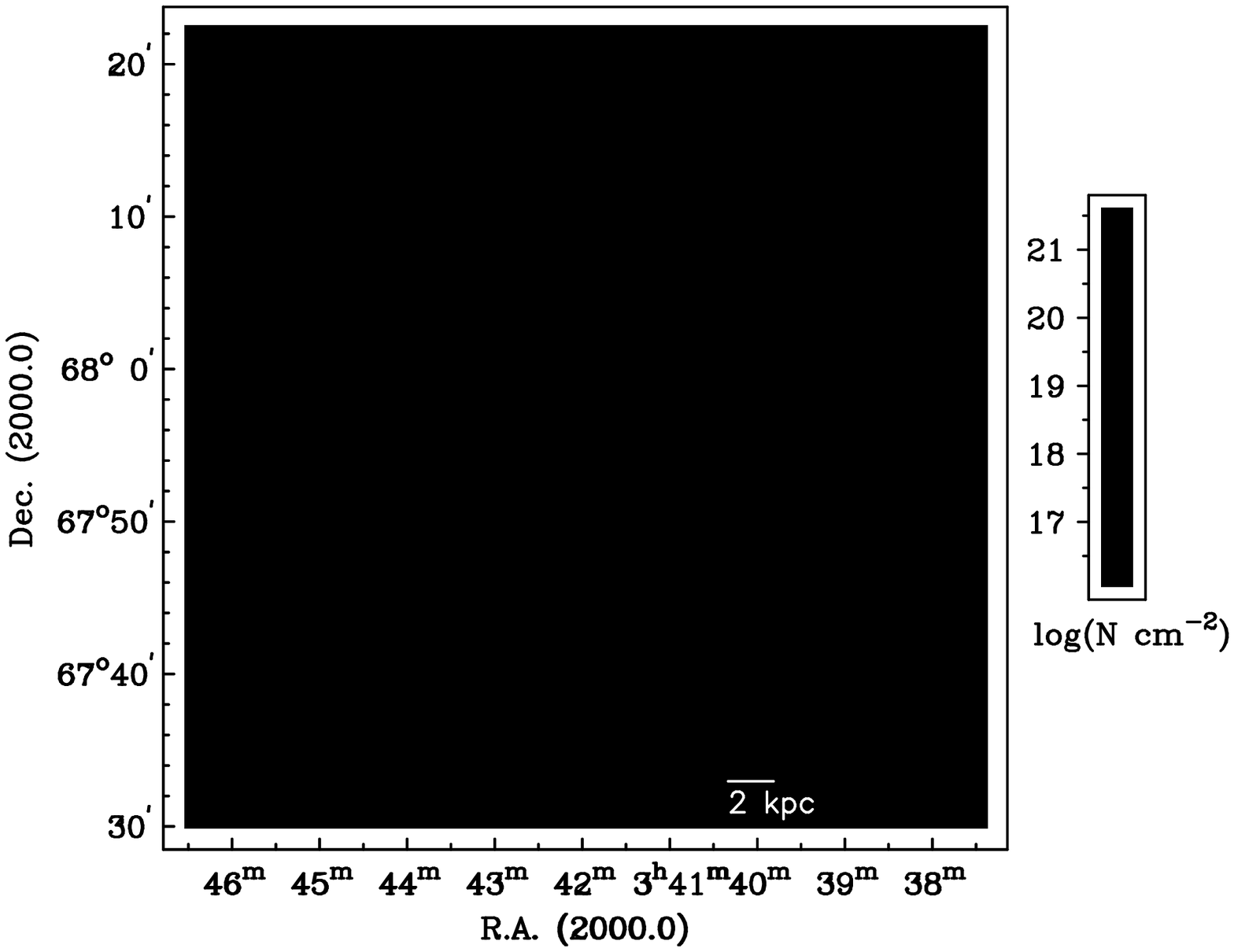}
\plottwo{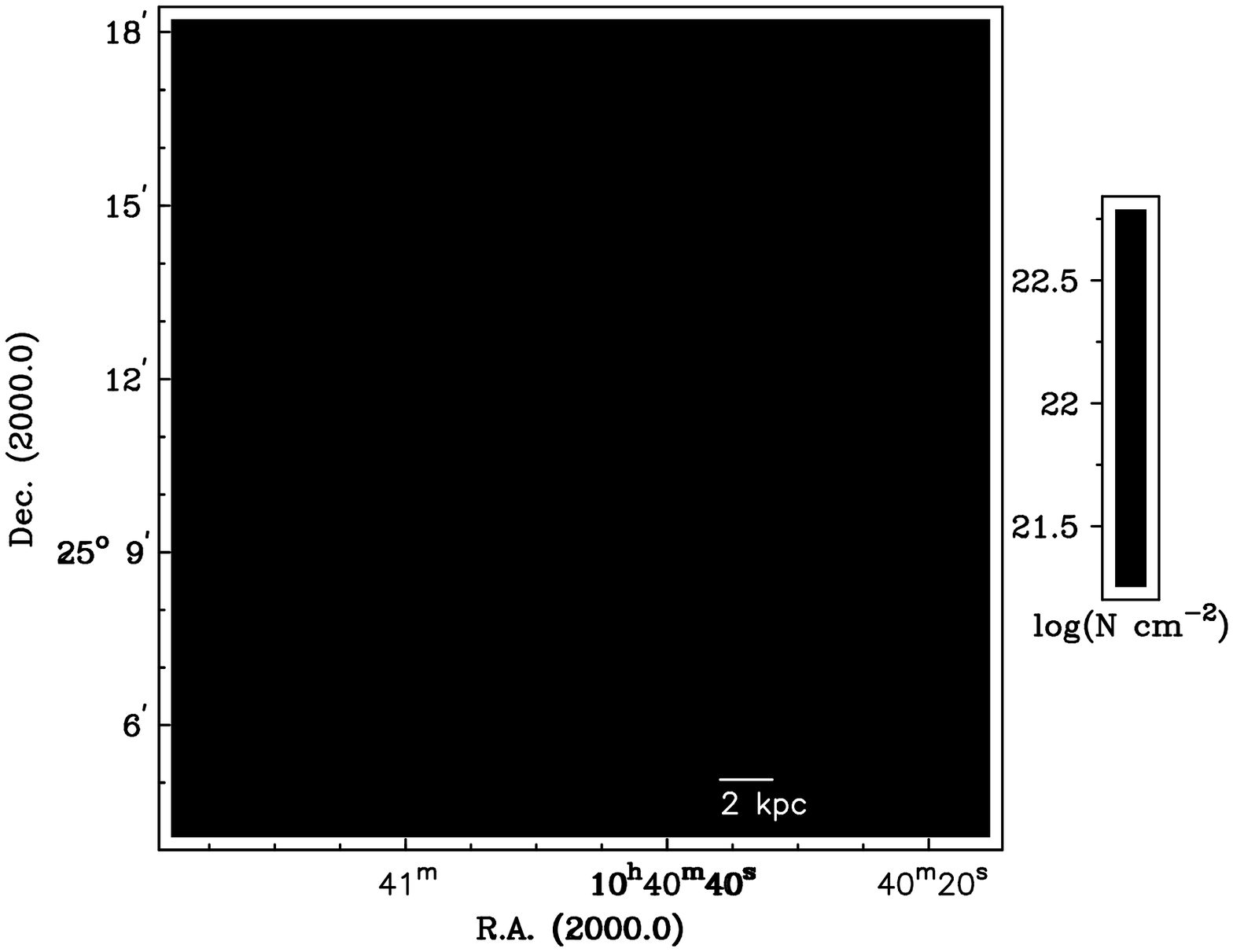}{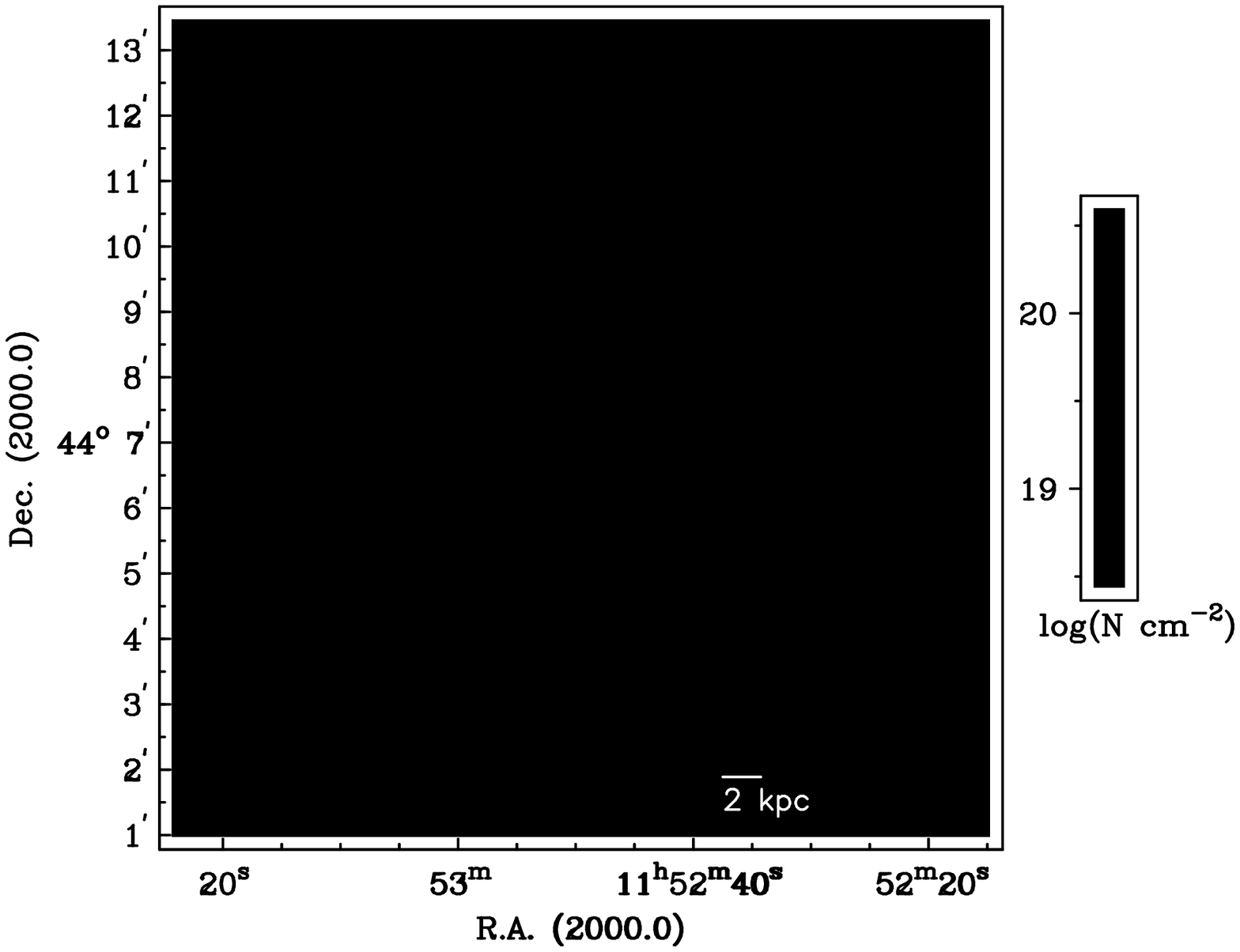}
 \end{center}
    \figcaption{Clockwise from top left: Zeroth moment maps for M101, IC 342, NGC 3938 (made from the combined CO and HI cubes) and NGC 3344 (made from the HI cube) showing the logarithm of the column density $N$ in units of cm$^{-2}$.  The horizontal bar near the bottom indicates the physical scale.  
\label{fig-mom0}}
\end{figure*}
\begin{figure*}
\begin{center}
\epsscale{1.}
\plottwo{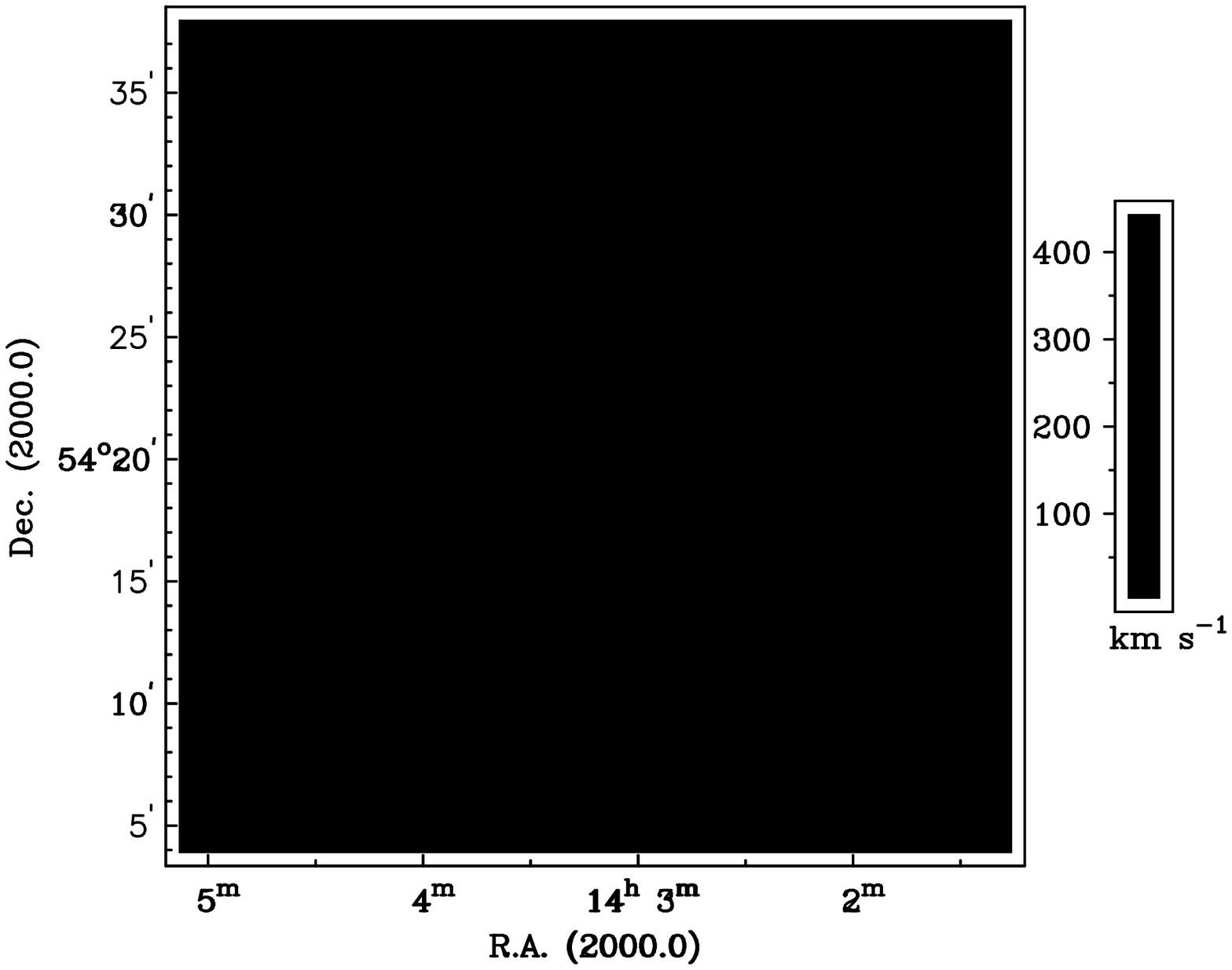}{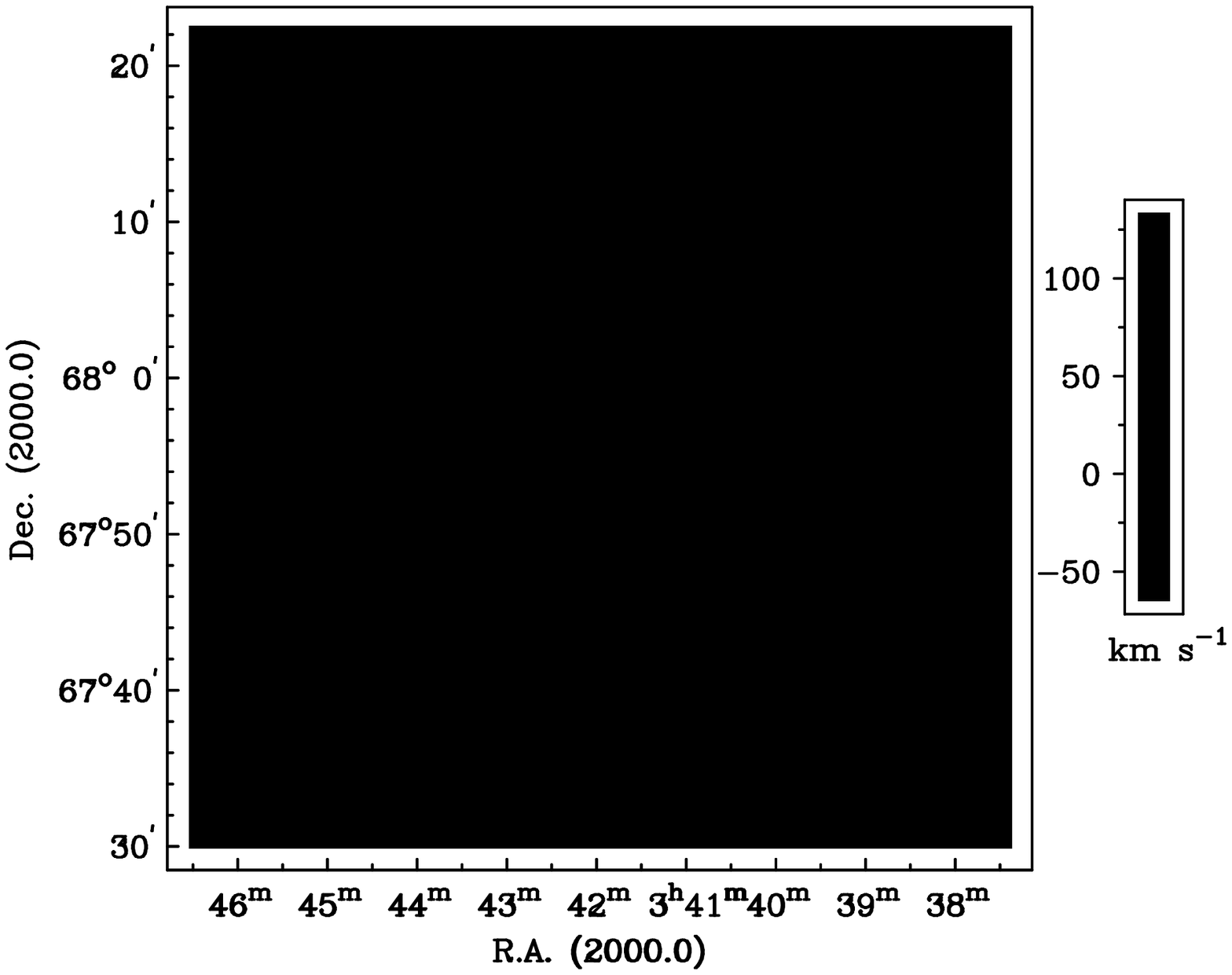}
\plottwo{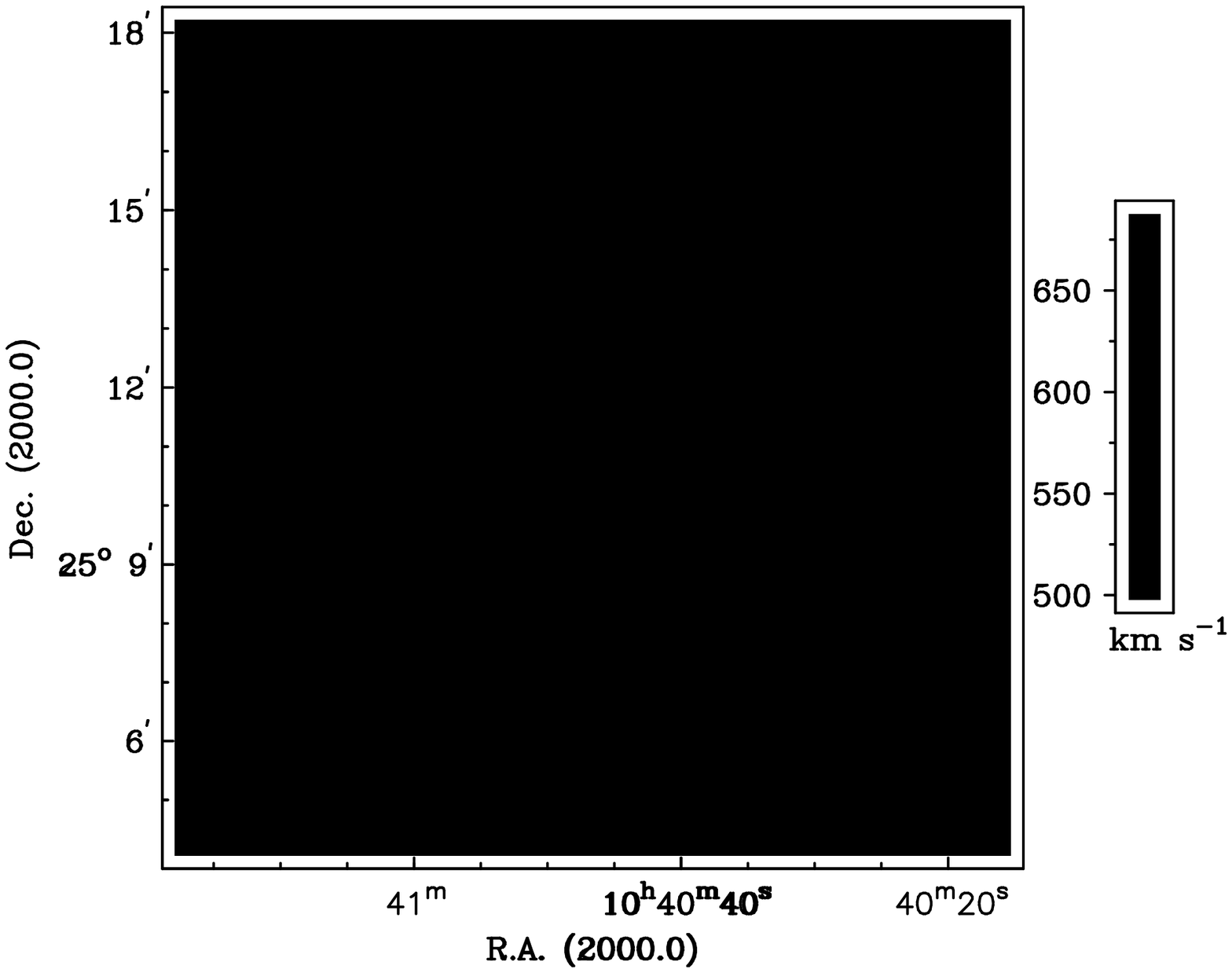}{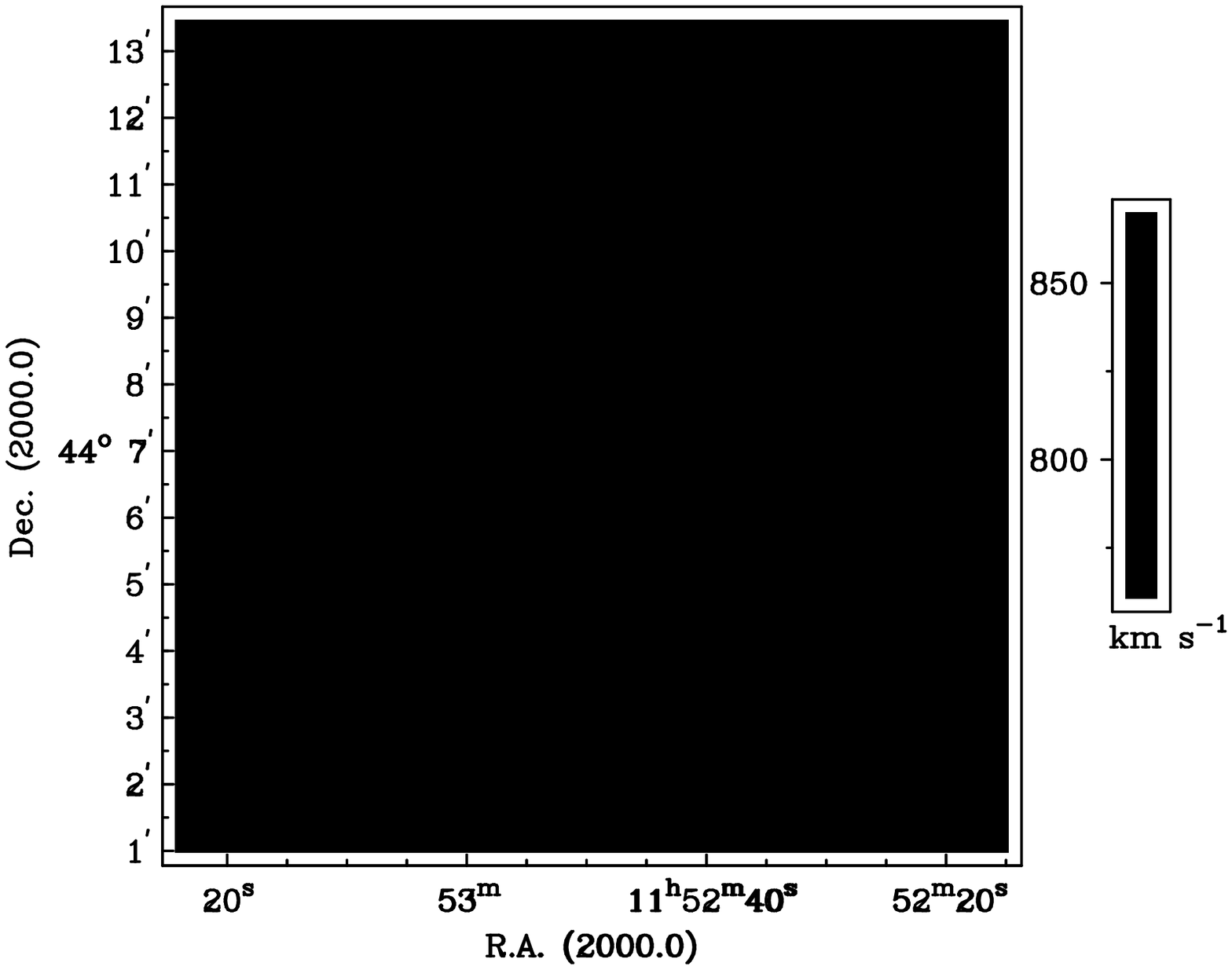}
 \end{center}
    \figcaption{Clockwise from top left: First moment maps for M101, NGC 3938, IC 342 (made from the combined CO and HI cubes) and NGC 3344 (made from the HI cube) in units of km s$^{-1}$.  
\label{fig-mom1}}
\end{figure*}
\begin{figure*}
\begin{center}
\epsscale{1.0}
\plottwo{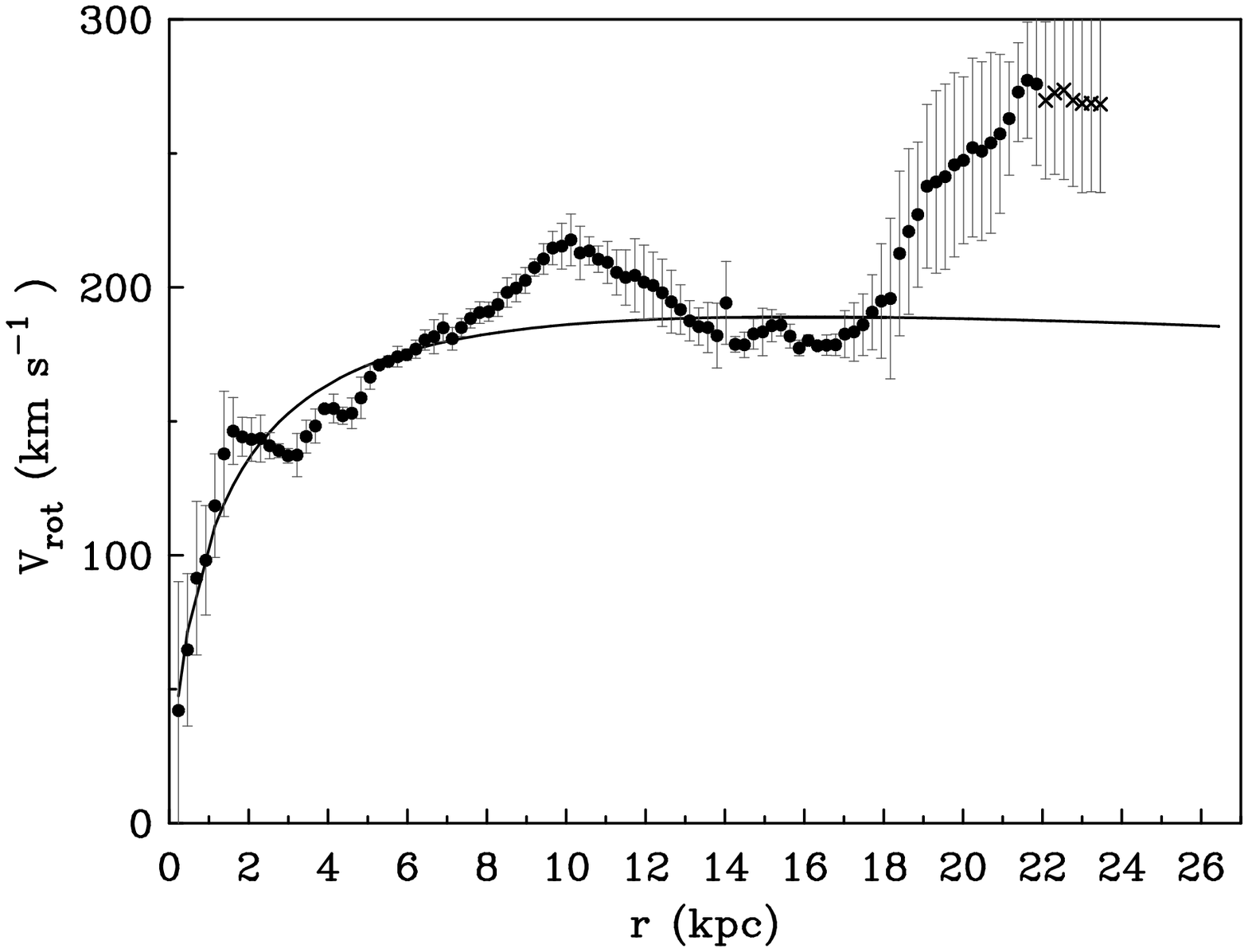}{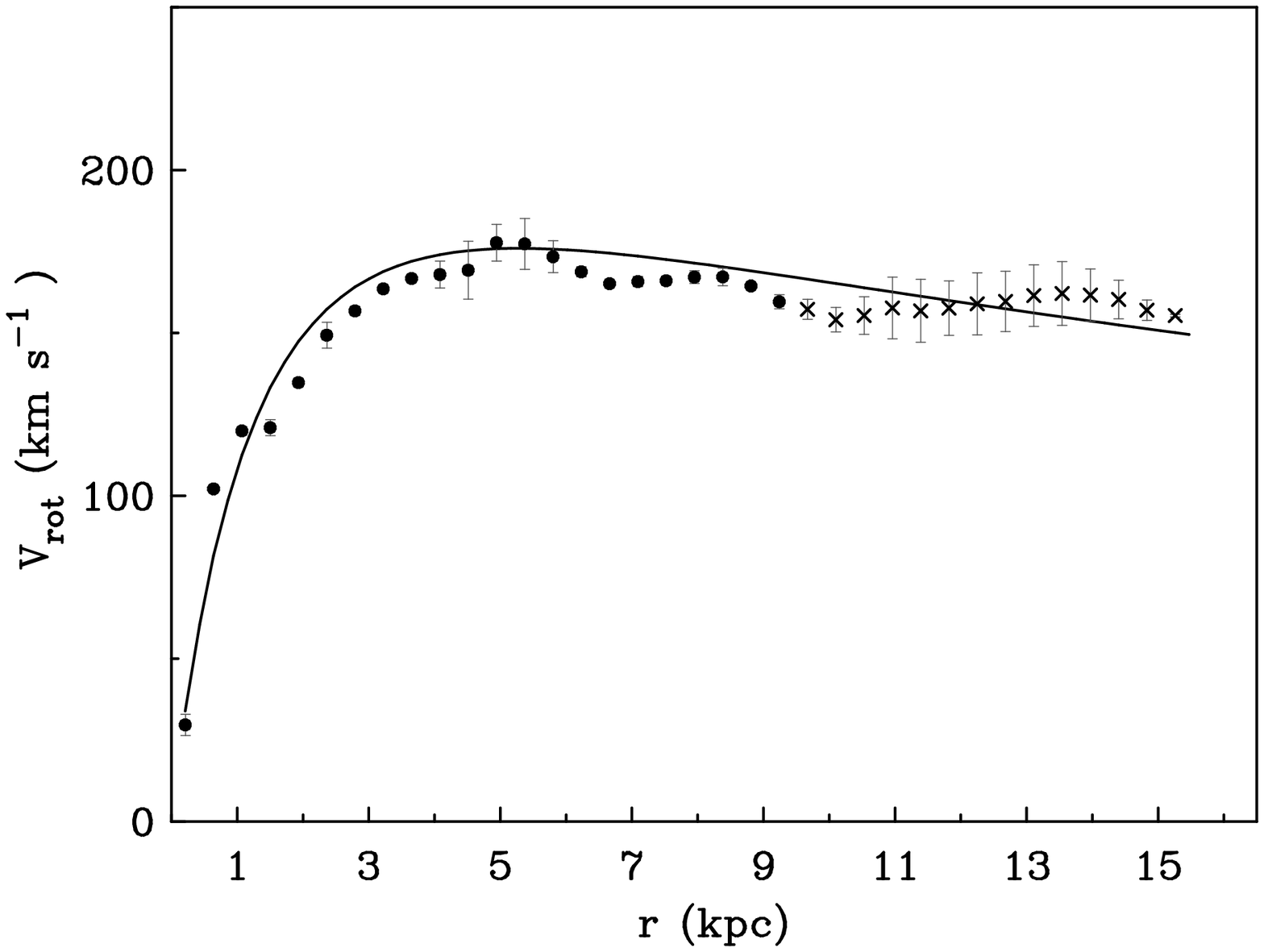}
\plottwo{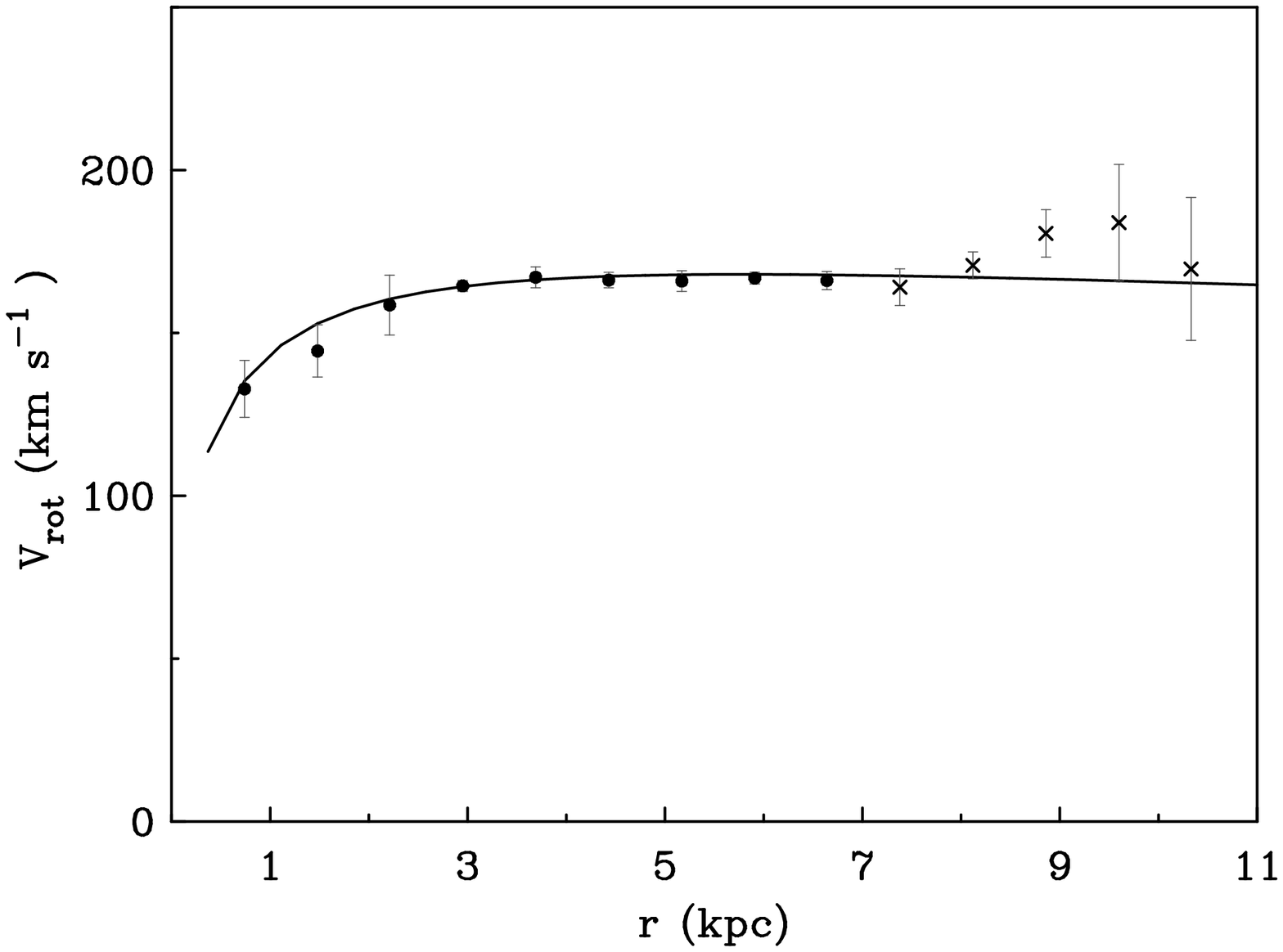}{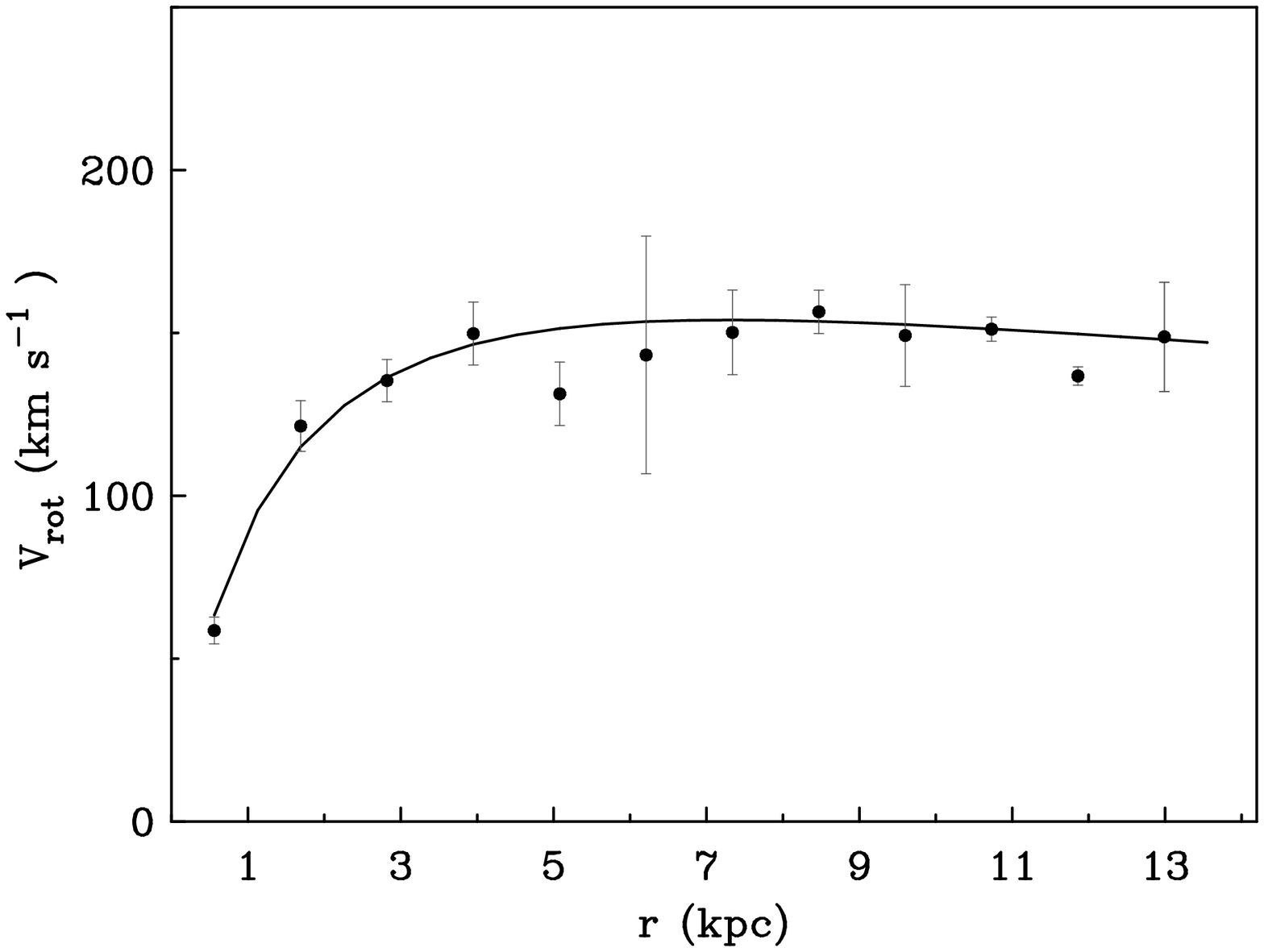}
 \end{center}
    \figcaption{Clockwise from top left: Rotational velocities for M101, IC 342, NGC 3938, and NGC 3344.  Values measured with tilted ring fits to the velocity fields shown in Figure \ref{fig-mom1} with all optimal parameters listed in Table \ref{tab-params} held fixed are shown as filled circles (crosses) in the unwarped (warped) region of the disk (as identified in $\S\S$ \ref{sec:results5457}-\ref{sec:results3344}).  The points are shown spaced at the resolution of each map with error bars (shown in dark gray) representing the average variation in the derived velocity from side to side (approaching and receding).  
The solid line in the plots for M101, IC 342 and NGC 3938 is the least-squares fit of the velocity model (Equation \ref{eq:vrot}) to the velocities measured in the unwarped portion of the disk (filled circles), with best-fit parameters: $V_{max}$=189 km s$^{-1}$, $r_{max}$=15.89 kpc, $n$=0.43 for M101; $V_{max}$=167 km s$^{-1}$, $r_{max}$=5.62 kpc, $n$=0.76 for IC 342; $V_{max}$=154 km s$^{-1}$, $r_{max}$=7.3 kpc, $n$=0.76 for NGC 3938; $V_{max}$=168 km s$^{-1}$, $r_{max}$=5.8 kpc, $n$=0.29 for NGC 3344. (For NGC 3344, to achieve an accurate fit these parameters had to be constrained a priori to a range that best reproduces the flatness for $r$$\gtrsim$3 kpc).  All solid lines are used to derive the smoothed angular rotation and resonance curves presented in $\S\S$ \ref{sec:results5457}-\ref{sec:results3344}.
\label{fig-vfits}}
\end{figure*}
In the following subsections we describe the four spiral galaxies analyzed in this paper, M101, IC 342, NGC 3938, and NGC 3344.  This small sample is not meant to be representative of variations in spiral structure with galaxy classification type, or to embody all possible theories of spiral structure.  These nearby galaxies were selected simply for their clear spiral structure and moderate inclination, as well as for the availability of observations (specifically, both CO and HI in three of the four cases) of sufficient sensitivity and resolution to allow structure to be resolved and the radial variation of $\Omega_p$ to be investigated; see Paper I.\\
\indent  With this sample we explore three scenarios for applying the TWR method to observations of the ISM.  We first consider both the molecular and atomic components of the ISM in M101, which is molecule-dominated over the detectable extent of the CO emission, and investigate whether there exists a relation between the speeds of the multiple structures apparent throughout the disk.  We then apply the method to two galaxies (IC 342 and NGC 3938) where the ISMs are dominated by neither the atomic or molecular gas phases in the central few kpc; in this case, considering the total column density is necessary for meeting the continuity requirement of the method.  Lastly, for NGC 3344, where the ISM is dominated by the atomic phase over the majority of the disk area (see $\S$\ref{sec:samp3344}), we analyze HI data alone, and with our TWR solution predict a pattern speed for the small bar.  \\
\indent  In three of the four cases, a warp-like distortion is evident in the HI distribution and kinematics, often with little further evidence of a strong spiral pattern.  
A warp in the outer disk, characterized by position and/or inclination angles that differ from those values in the disk interior, violates the TW assumption that the disk is flat \citep{tw84}.  In implementing the regularized TWR method we therefore adopt the procedure advocated in Paper I (see the beginning of $\S$ \ref{sec:results}), which accommodates for the presence of information about the warp in the TWR quadrature by excluding it from regularized solutions.  \\
\indent Zeroth and first moment maps for each galaxy are shown in Figures \ref{fig-mom0} and \ref{fig-mom1}.  For M101, IC 342 and NGC 3938, these maps were generated from the combined CO and HI cubes, which were regridded to a uniform channel width and co-added in units of particles cm$^{-2}$.  Prior to combination, the CO cubes were smoothed to the resolution of the HI and CO intensities converted to molecular column densities assuming a constant CO-to-$H_2$ conversion factor $X$=2.0$\times$10$^{20}$cm$^{-2}[$K km s$^{-1}]^{-1}$, the Galactic mean value (e.g. \citealt{hunter}).  (Subsequent variations in the $X$-factor are considered when applicable.)  \\
\indent Tilted ring fits to the velocity field of the four galaxies were performed using the GIPSY \citep{gipsy} task ROTCUR in order to derive the kinematic parameters used in the TWR calculation.  For these fits, the systemic velocity is initially fixed to the value given in the literature, while the kinematic center, position angle (PA), and inclination are allowed to vary.  (The value and the errors on $V_{sys}$ were subsequently determined by fitting with all other best-fit parameters held fixed.)  The values we find are in good agreement with those nominally adopted for each galaxy and are listed in Table \ref{tab-params}.  \\
\indent For the purposes of identifying resonances (corotation, inner and outer Lindblad), we also derived the circular velocity in each ring with the best-fit kinematic parameters held fixed.  In each galaxy these velocities are modeled in the unwarped portion of the disk with least-squares fits to the standard three-parameter approximation (i.e. \citealt{fabergall}) 
\begin{equation}
V_{rot}(r)=\frac{V_{max}(r/r_{max})}{\left(1/3+2/3(r/r_{max})^n\right)^{3/2n}} \label{eq:vrot}
\end{equation}
as shown in Figure \ref{fig-vfits}.  The errors bars on each measured velocity there reflect the average deviation from this value on either the approaching or receding side, which together we find better represent the uncertainty in the rotation curve than do the formal errors returned by ROTCUR.  With the fitted velocities, we then generate a set of smooth curves for $\Omega$, $\Omega$$\pm$$\kappa$/2, and $\Omega$$\pm$$\kappa$/4; these curves are intended to reduce the impact of non-axisymmetric motions (e.g. spiral streaming) on our resonance identifications and were invoked without regard to specific mass component characterization.  
\begin{table*}
\begin{center}
\caption{Parameters used in the TWR calculation.\label{tab-params}}
\begin{tabular}{rcccc}
\tableline\tableline

 Parameter& M101&IC 342 &NGC 3938 & NGC 3344\\
\tableline
 Dynamical Center RA ($\alpha$) (J2000)&14$^h$3$^m$13$^s$.13&3$^h$46$^m$48$^s$.4&11$^h$52$^m$49$^s$.8&10$^h$43$^m$31$^s$.5\\
 Dynamical Center DEC ($\delta$) (J2000)&54$\degree$20'56''&68$\degree$5'47''.8&44$\degree$7'11''.7&24$\degree$55'18''.3\\
 Distance (Mpc) &7.4 &2.0&11.3 &6.9\\
 Systemic Velocity (V$_{sys}$, km s$^{-1}$)&244$\pm$8&30$\pm$2&809$\pm$3&586$\pm$3\\
 Position Angle($\degree$) &42$\pm$3&42$\pm$3 &21$\pm$2&155$\pm$2\\
 Inclination ($\degree$) &21$\pm$6&31$\pm$5&14$\pm$3&25$\pm$4\\
\tableline
\end{tabular}
\tablecomments{Entries for the dynamical center, inclination and position angle for each galaxy were derived with a tilted ring analysis of the first moment of the data cube using the GIPSY task ROTCUR.  Optimal values for the distance originate with references cited in $\S\S$ \ref{sec:samp5457}-\ref{sec:samp3344}. }
\end{center}
\end{table*} 
\subsection{\label{sec:samp5457}M101}
This SABcd galaxy (D=7.4 Mpc; \citealt{jurcevic}) is an excellent candidate for analysis with the TWR method.  
The molecular gas filling the central 3' hole in the HI emission features a bar discovered by \citet{ksw}.  The bar is a clear candidate-driver of the spiral structure that appears to emanate from the bar region.  At the outer radii, tidal interaction with companion galaxies (e.g. NGC 5474 and NGC 5477; \citealt{huch}, \citealt{waller}) is thought to be responsible for the lop-sidedness and distortion in the HI distribution, and may also be the source of the spiral \citep{waller}.  
The high resolution ($\sim$7'') of the total H$_2$+HI maps constructed from the BIMA SONG CO data and THINGS HI data (with ROBUST weighting scheme; \citealt{walter}) allows us to apply the TWR method with exceptional leverage on the radial dependence of the pattern speed throughout the disk.  We aim to derive the bar and spiral pattern speeds, and identify multiple spiral modes and spiral winding, if present.\\
\indent As considered in the upcoming discussion ($\S$ \ref{sec:results5457}), we note that the rotational velocities between $r$$\sim$7 and 14 kpc and at radii $r$$\gtrsim$19 kpc in the top right of Figure \ref{fig-vfits} are not well fit here.  Also, while the rise in the latter zone appears on both the approaching and receding sides, a $\sim$75 km s$^{-1}$ asymmetry exists therein.  \\
\indent In addition, as demonstrated by \citet{wb} assuming $X$=1.8$\times$10$^{20}$cm$^{-2}$$[$K km s$^{-1}]^{-1}$ (slightly lower than our adopted value), and using the BIMA SONG data we analyze here, the ISM is molecule-dominated over the extent of the CO emission.  However, the disk is known to sustain a metallicity gradient, which, for a linear dependence of $X$ on metallicity (e.g. \citealt{boselli}), could imply variation in $X$ with radius.  So, while a constant offset should have no implications for TWR solutions, we also consider the effect of variation in $X$ outward from the center.
\subsection{\label{sec:samp342}IC 342}
For this Seyfert SABcd galaxy (D=2.0 Mpc; \citealt{crosthwaiteHI}), we base our analysis on the BIMA SONG CO data cube and the 38'' VLA HI observations first published by \citet{crosthwaiteHI}.  
As in M101, the CO fills the central hole in HI distribution, but here the atomic gas makes a significant contribution to the total gas density before the edge of the CO emission.  A thorough discussion of the HI-CO overlap and the features (a bar, a two-armed spiral, and a four-armed spiral) in the two gaseous components can be found in the study of \citet{crosthwaiteCO}.  
\subsection{\label{sec:samp3938}NGC 3938}
\indent NGC 3938 (D=11.3 Mpc; \citealt{jv}) is a nearly face-on, late type (SAc) galaxy exhibiting a central two-armed spiral that branches into multi-armed structure in the optical.  The two strong spiral arms are evident in the molecular gas, which reaches a surface density comparable to that of the HI inside the edge of the CO emission.  But where the stellar disk exhibits flocculent, but clear, spiral structure (with as many as six arms; \citealt{eem}), the HI disk is characterized by less well-organized, irregular structure.  The combination of archival WSRT HI data with BIMA SONG CO observations establishes a total H$_2$+HI kinematic tracer at roughly 20'' resolution that extends to just beyond the optical extent of this galaxy \citep{vs}.  
\subsection{\label{sec:samp3344}NGC 3344}
The application of the model-independent TWR method to this HI-dominated, isolated SABbc galaxy (D=7.4 Mpc; \citealt{vm}) should, in principle, clearly establish the relation of the ring-like morphological features at $r$=1 kpc and $r$=7 kpc, identified by \citet{vm} in optical images, to the large-scale patterns present in the disk (and their resonances).  The outer ring, in particular, although thought unlikely to be related to the small bar (not observed in the gas) with $a_B$$\sim$0.7 kpc \citep{vm}, has not been otherwise conclusively related to the spiral.  In search, as well, of spiral winding and multiple spiral modes, here we analyze 20'' resolution archival WSRT HI data where the dominant two-armed spiral and the outer ring are clear, in addition to the outer lopsided region exhibiting a twist in the isovels \citep{vs}.  The zone of the bar and the inner ring, which falls within the central 26'' where there is little 21-cm emission, corresponds to less than
two resolution elements.  So although the ring is resolved in CO \citep{regan}, and the peak H$_2$ column density \citep{helfer} there exceeds the HI, we do not consider the contribution of the molecular gas here. 
\section{Results}
\subsection{\label{sec:results}Applying the TWR Method}
\indent We apply the regularized TWR method as in Paper I.  For each galaxy we consider several smoothed, testable models for $\Omega_p(r)$.  These models vary as polynomials (order $n$$\lesssim$2) designated into at most three distinct radial zones.  
Where a priori evidence suggests that there is little information from a strong pattern beyond a certain radius, or that the TW assumptions are otherwise violated by the presence of a warp, for example, our models also include the parameterization of a cut radius, $r_c$, beyond which all bins are calculated without regularization (i.e. the functional form is unconstrained).  These models, with $r_c$ marking the end of the dominant structure, have been demonstrated to sufficiently separate the compromised zones in the disk from regions where information about patterns can be reliably extracted in the TWR calculation (Papers I and II).  All transitions $r_t$ between distinct zones and all cut radii, where present, are treated as free parameters. \\ 
\indent For each model, the two numerical solutions for $\Omega_p(r)$ from either side of the galaxy (y$>$0 and y$<$0; \citealt{mrm}) are averaged to construct a single, global model solution.  Each model solution is then judged based on a simple reduced $\chi^2$ statistic, with the best-fit solution corresponding to the $\chi^2_{\nu}$-minimum in the full parameter space.  \\
\indent As in Paper II, the random, measurement errors used in the regularized calculation, and with which we judge the best-fit solution, reflect uncertainty arising with the chosen flux cut-off in the first moment maps.  The systematic errors on each measurement represent uncertainty in the PA, which is the dominant source of error in TW and TWR estimates (\citealt{debPA}; Paper I), roughly 20\% for $\delta_{PA}$=3$\degree$.  But, here we report these as a dispersion on each measured value, rather than present individual solutions for each PA; this is possible here since, unlike in M51 (Paper II), we find no meaningful evidence that the form of the model associated with the best-fit solution for any of the galaxies in the current sample varies from PA to PA.  Also, unless otherwise noted, errors due to uncertainty in, e.g., the inclination angle are generally smaller and are not reported; these prove to be of little consequence to the accurate placement of radial bins defined in the quadrature (as suggested in Paper I).  The additional change introduced in the measurements $\Omega_p$ through a change in $\sin{i}$ is furthermore shared by $\Omega$ and $\kappa$, and so our resonance identifications, in particular, should not be effected by error in the inclination to first order.  A thorough account of our methodology can be found in Paper I (and references therein).
\subsection{\label{sec:results5457}M101}
\begin{figure}
\begin{center}
 \leavevmode
\plotone{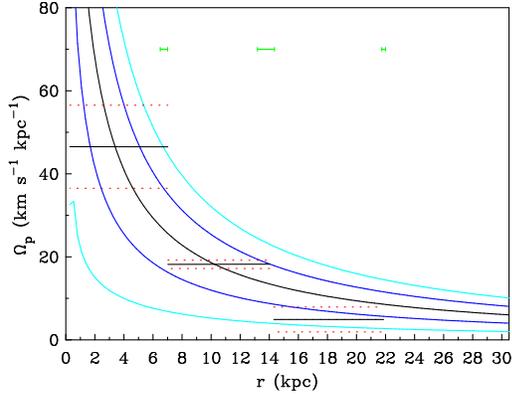}
\end{center}
\figcaption{
Best-fit regularized solution for M101 with $r_c$=21.9$\pm$0.43 kpc for PA=42$\degree$$\pm$3$\degree$.  For this solution, bins exterior to $r_C$ (not shown) have been calculated without regularization.  Dashed red lines represent the dispersion in solutions derived with a three-pattern speed model at PA=39$\degree$ and 45$\degree$.  Horizontal error bars represent the dispersion in $r_{t,1}$, $r_{t,2}$ and $r_c$ from PA to PA.  The innermost speed corresponds to $\Omega_{p,1}$=47$\pm$10 $\kmskpc$ out to $r_{t,1}$=6.7$\pm$0.25 kpc, followed by $\Omega_{p,2}$$\sim$18$\pm$1 $\kmskpc$ out to $r_{t,2}$=13.8$\pm$0.58 kpc and $\Omega_{p,3}$=5$\pm$3 $\kmskpc$ out to $r_C$.  Curves for $\Omega$, $\Omega$$\pm$$\kappa$/2 and $\Omega$$\pm$$\kappa$/4 (see text) are shown in black, cyan and blue. \label{fig-5457twr}}
\end{figure}
\begin{figure}
\begin{center}
\epsscale{1.0}
\plotone{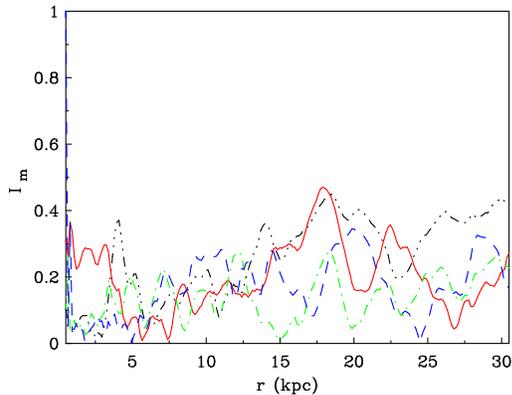}
 \end{center}
    \figcaption{Fourier power spectrum of the zeroth moment map shown in the top left of Figure \ref{fig-mom0}.  Modes up to $m$=4 are plotted as a function of radius with lines for $m$=1 in black dash-dot-dot-dot, $m$=2 in red solid, $m$=3 in green dash-dot, and $m$=4 in blue dash.
\label{fig-5457fourier}}
\end{figure}
\begin{figure}
\begin{center}
\epsscale{1.0}
\plotone{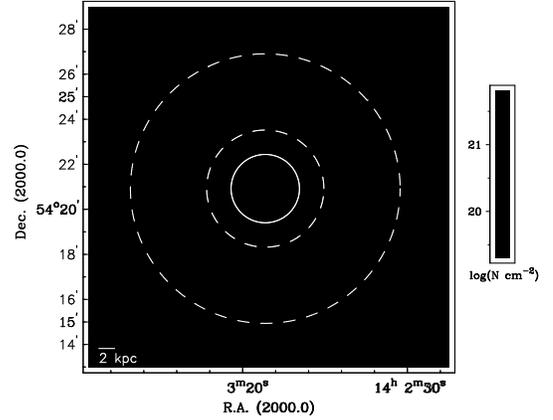}
 \end{center}
    \figcaption{Total H$_2$+HI surface density in M101 highlighting the structure inside $\sim$18 kpc.  The transitions $r_{t,1}$ and $r_{t,2}$ marking the extents of the first and second pattern speeds in our solution are shown as dashed white circles.  The corotation radius $r_{CR}$=3.5 of the innermost speed is shown as a solid white circle.  The horizontal bar near the bottom left indicates the physical scale.  
\label{fig-5457zoom}}
\end{figure}
\indent We apply the TWR calculation with radial bin width $\Delta r$=7''=0.27 kpc (D=7.4 Mpc), the resolution of the combined cube.  Together with the position of the outermost slice on each side, $\vert y\vert$=30.4 $\cos{i}$ kpc, this establishes the extent of integration along each slice, equipping solutions with 113 bins in total.  The best-fit solution given a PA uncertainty $\pm$3$\degree$ is plotted in Figure \ref{fig-5457twr}.  There shown, also, are the rotation and resonance curves derived as described in $\S$\ref{sec:sample}.  It should be noted that a $\sim$75 km s$^{-1}$ asymmetry in the rotational velocities from the approaching and receding sides exists at radii $r$$\geq$20 kpc (e.g. \citealt{kamphuis}; \citealt{jogv1}).   \\
\indent The outermost portion of the disk is effectively removed from the solution with the parameterization of a cut radius $r_c$=21.9$\pm$0.43 kpc.  This radius, identified at the minimum of the $\chi^2$, is comparable to the location where the disk becomes visibly distorted.  We find that the outer distortion/lop-sidedness clear in the surface density is well characterized by the predominance of an $m$=1 asymmetry beyond $r$$\sim$20 kpc in the Fourier decomposition (Figure \ref{fig-5457fourier}) and also matched by a warp in the outer velocity field; in addition to the rotation curve asymmetry, with our ROTCUR analysis we find that the PA and inclination of fitted rings beyond this radius vary significantly from the nominal values established in rings interior. (We note that this strong variation in the PA is found to start at a much larger radius than identified by \citealt{rdh}.) Inclination variation, in particular, is a violation of the TW/TWR assumptions and so we argue that, by excluding the bins covering the outer disk from models for $\Omega_p(r)$, the remaining, inner regularized bins are better equipped to reproduce the true pattern speed. (Note that with this cut radius a possible $m$=1 mode describing the outer, lop-sided portion of the disk is ignored).\\ 
\indent In this case, the most conspicuous aspect of the solution is the pair of transitions at $r_{t,1}$=6.7$\pm$0.25 kpc and $r_{t,2}$=13.8$\pm$0.58 kpc (marked in Figure \ref{fig-5457zoom}) between three distinct pattern speeds, $\Omega_{p,1}$=47$\pm$10 $\kmskpc$ $\Omega_{p,2}$=18$\pm$1 $\kmskpc$ and $\Omega_{p,3}$=5$\pm$3 $\kmskpc$ out to $r_c$.  As discussed below, the innermost, constant speed is nearly identical to the value $\Omega_p$=48$\pm$8 $\kmskpc$ measured from the CO maps, alone (see $\S$ \ref{sec:m101inner}), while the second pattern speed is consistent with the value generally upheld in the literature ($\Omega_p$$\sim$19 $\kmskpc$ for $r_{CR}$$\sim$12 kpc, e.g. \citealt{waller}; but see the end of this section).  \\
\indent The quality of our measurement 
is especially clear in the precision with which the transitions between the three pattern speeds are determined.  The transition $r_{t,1}$, for example, shows less than 5\% variation from PA to PA.   
This transition coincides with a marked decrease in $m$=2 power (Figure \ref{fig-5457fourier}) and occurs well past the edge of the CO emission.  We can therefore recognize that the inner speed describes the molecular bar and the two-armed spiral manifest in the CO and weakly traceable in HI, which \citet{eem} identify in the B-band out to $r$$\sim$6 kpc (estimated from their placement of the inner 4:1 circle in Figure 1, Plate 14, rather than with the discrepant value listed in Table 3 there).  (Yet, we also note a four-armed appearance to the structure where spur-like features associated with either arm emerge, clear in CO emission and in the B-band). \\ 
\indent This result, namely that structure in HI shares the same pattern speed as the structure interior (exhibited solely by the molecular gas), admits a much stronger conclusion than can be drawn from the CO, alone.  Our rotation and resonance curves 
 indicate that the inner pattern ends near its OLR ($r$$\sim$7 kpc; or possibly the outer 4:1 resonance, within the errors) which would not be evident from the application of the TWR method to CO data covering $r$$\lesssim$3.5 kpc (corresponding to the central zone in Figures \ref{fig-mom0} and \ref{fig-5457zoom} with column density $\gtrsim$10$^{20.6}$ cm$^{-2}$).  This finding is consistent with the theoretical expectation for the forbidden propagation of spiral density waves beyond this resonance, which observations corroborate (see \citealt{elm89}).  \\
\indent Likewise, corotation for this speed occurs at $r$=3.5 kpc, very near the transition between molecular and atomic gas (see Figure \ref{fig-5457zoom}).  With its lack of clear structure and relatively low surface density, this location in the gaseous disk seems consistent with an expected depopulation near CR owing to opposing torques (whereby gas is driven inward between ILR and CR and outward between CR and OLR).  (This impression, however, may be subject to the sensitivity of the CO map and the assumed $X$-factor.) \\
\indent Interior to corotation, our determination of the disk angular rotation becomes less certain at a disadvantage to definitive resonance identifications.  Nevertheless, the two-armed spiral exhibited near $r$$\sim$2 kpc by the molecular gas--and which appears almost ring-like--is arguably located near the inner 4:1 resonance kpc, or UHR, where gas can accumulate on its path inward from corotation to ILR \citep{bc}.  
In addition, as will be discussed in more detail in $\S$\ref{sec:m101inner}, the central concentration of gas in the form of the molecular bar (with length $\sim$1 kpc) appears to lie well within the CR.  This is also true of the stellar bar with $a_b$$\sim$2 kpc \citep{ksw}, which suggests a scenario quite different from observational findings in favor almost exclusively of fast stellar bars (with 1$\lesssim$$r_{CR}$/a$_b$$\lesssim$1.4; \citealt{corsini}), at least in early-type barred galaxies.  Instead, the stellar bar here seems reminiscent of the slow bars found in the simulations of \citet{ce} and \citet{rsl}.  \\
\indent 
As for the pair of pattern speeds $\Omega_{p,2}$ and $\Omega_{p,3}$ outside $r_{t,1}$=6.7$\pm$0.25 kpc, we can similarly use the curves in Figure \ref{fig-5457twr} to interpret their radial domains in terms of resonances.  This is less straightforward, however, primarily because of our uncertainty in the poorly-fit rotation curve here, as well as the complex nature of the spiral structure visible in the HI surface density, which is characterized by power in several Fourier modes (see Figure \ref{fig-5457fourier} and optically, \citealt{eem}). Notably, between $r$$\sim$6 kpc and 13 kpc, the structure visible in HI appears four-armed (see the top left panel in Figure \ref{fig-mom0} and Figure \ref{fig-5457fourier}).  Meanwhile, near $r$=13 kpc a two-armed pattern starts to predominate again, as indicated in Figure \ref{fig-mom0} and by Figure \ref{fig-5457fourier} where the power in the $m$=4 mode decreases relative to the $m$=2 mode.  \\
\indent In this case, the transition $r_{t,1}$ in Figure \ref{fig-5457twr} would seem to associate the second speed $\Omega_{p,2}$ with the four-armed spiral, here found to extend between the inner and outer 4:1 resonances at $r$$\sim$6.8 kpc and $r$$\sim$14 kpc. (This is unchanged even with rotation curve models where the rise and fall in velocities between $r$=8 and 14 kpc in Figure \ref{fig-vfits} are more closely (yet coarsely) fit, although the end of $\Omega_{p,2}$ more nearly approaches $\Omega$-$\kappa$/2).  The four-armed spiral in ESO 566-24 likewise extends between its inner and outer 4:1 resonances \citep{rsb}.  Our finding is also roughly consistent with the identification made by \citeauthor{eem} (\citeyear{eem}; in Figure 1, Panel 14 there), namely that the zone dominated by the four-armed spiral is bounded by the inner 4:1 circle with $r$$\sim$6 kpc and the CR circle with $r$$\sim$12 kpc.  (We would argue, however, that the end of this zone occurs past the CR, which we find near $r$=10 kpc, depending on the rotation curve.)  \\
\indent This is a scenario moreover favored by our TWR solution in that the transition to the four-armed spiral occurs at a resonance overlap; where $\Omega_{p,1}$ ends at OLR, within the uncertainties, $\Omega_{p,2}$ begins near its inner 4:1 resonance at $r$$\sim$7 kpc.   
Such an alignment of resonances, although not conclusively associated with mode-coupling (i.e. as opposed to CR-ILR overlaps identified by \citealt{masset2}), has been identified in the simulations of \citet{rs} and \cite{debatSim}, as well as in the grand-design spiral M51 through application of the TWR method (Paper II).  Presumably, this overlap is characteristic of a physical mechanism by which spiral structure can be sustained over a large span in radius \citep{rs}.  \\
\indent The transition $r_{t,2}$ between $\Omega_{p,2}$ and $\Omega_{p,3}$ may also be accompanied by resonance overlap: the outermost speed, which spans a radial domain well-matched to that of the two-armed spiral within 12$\lesssim$$r$$\lesssim$20 kpc (marked by the increase in $m$=2 power in Figure \ref{fig-5457fourier}), could begin at either the ILR or inner 4:1 resonance.  Large fractional errors on $\Omega_{p,3}$, of course, make this identification particularly vague.  Likewise, the end of this pattern, as designated by the cut radius, can be only indefinitely related to realistic resonances (e.g. CR; at least with our current determination of the angular rotation), although it is just beyond the end of the outer two-armed spiral identified by \citet{eem} near $r$=19 kpc (again, according to their Figure 1, Panel 14, rather than Table 3).  In addition, whether or not the resemblance to $\Omega$-$\kappa$/2, $\Omega$-$\kappa$/4 or $\Omega$ suggested in Figure \ref{fig-5457twr} is significant remains unclear at this point, especially in light of the nearly linear rise (and the asymmetry) in the rotation velocities outside $r$$\sim$19 kpc (not well modeled here; see Figure \ref{fig-vfits}) suggesting that the angular rotation curve may flatten out near this radius.  The values of $\Omega$-$\kappa$/2 and $\Omega$-$\kappa$/4 would be lower (and $\Omega$+$\kappa$/2 and $\Omega$+$\kappa$/4 higher) than in Figure \ref{fig-5457twr}, in which case $\Omega_{p,3}$ may coincide with $\Omega$-$\kappa$/4.  \\
\indent Despite our uncertainty in the outermost speed, overall, the TWR solution presents compelling evidence for extensive spiral structure described by multiple pattern speeds that are moreover related by their overlap at resonance.  
Given that this galaxy is tidally interacting  (e.g. as argued by \citealt{ideta}), our pattern speed solution may be suggestive of the scenario explored by \citet{salo} in simulations of M51.  Those authors find that waves of higher and higher pattern speeds are excited as tidally-induced waves near $\Omega$-$\kappa$/2 propagate inward.  
But considering that resonance overlap is not a feature in those simulations, our finding of a link between the speeds in M101 (and in M51; Paper II)--similar to the overlaps so far demonstrated between bars and spirals in simulation (e.g. \citealt{masset2}; \citealt{rs})--may imply that at least the inner disk is dominated by internal, rather than external, effects.
The speeds and structure within $r$$<$$r_c$$\sim$22 kpc, however, have yet to be related to the asymmetric remainder of the disk.  If the outer warp is a lop-sided, $m$=1 mode, its pattern speed (not measured here) may be determined by, and reveal clues to, tidal encounter with the companion galaxies of M101.  
\subsubsection{\label{sec:m101inner}The inner pattern speed in M101}
\begin{figure}
\begin{center}
\epsscale{1.0}
\plotone{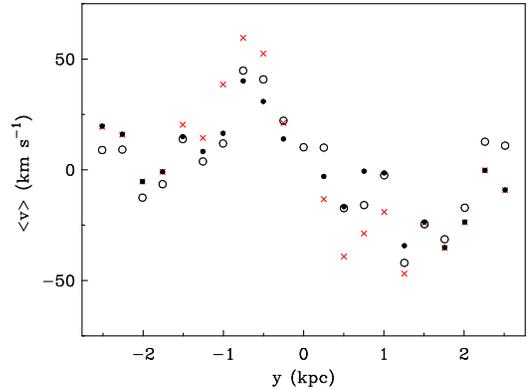}
 \end{center}
    \figcaption{Plot of solution-reproduced (dots and crosses) and actual (open circles) integrals $<$$v$$>_i$=$b_i$/$\int \Sigma dx$ (see Paper II) as a function of slice position $y$ at PA=42$\degree$ calculated from BIMA SONG CO data for M101.  The values associated with the best-fit single pattern speed solution (black dots) are plotted along with those of the two mock pattern speeds, $\Omega_{p,1}$=115 $\kmskpc$ out to $r_t$=1.1 kpc and $\Omega_{p,2}$=48 $\kmskpc$ throughout the remainder of CO-traced disk (red crosses).
\label{fig-5457vslice}}
\end{figure}

\indent As stated previously, the inner disk appears to sustain only a single constant pattern speed, $\Omega_{p,1}$=47$\pm$10 $\kmskpc$.  This may be surprising, given the distinct bar pattern visible in the CO emission, which we might expect to end near CR and rotate with a pattern speed distinct from that of the spiral pattern immediately exterior.  In order to determine whether our TWR solution is an accurate representation of the true speed, in this section we discuss the evidence in favor of only a single pattern speed in this zone, consider whether the measurement is an artifact of variation in the $X$-factor, and make comparisons to findings in the literature.\\
\indent For this investigation we consider the CO data alone. As might be expected given the radial domains of the two pattern speeds, we find that solutions calculated from the CO and HI maps, individually, supply pattern speed estimates nearly identical to those in either the inner or outer zones of the solution in Figure \ref{fig-5457twr}; the TWR result from the CO maps, $\Omega_p$=48$\pm$8 $\kmskpc$, is within the errors of the inner speed measured from the combined maps.  Furthermore, the contribution made by the bar and inner spiral pattern in a given slice can be more easily identified when slices do not also contain information from throughout zone of the HI.  (Note that, without the need for a cut radius, TWR solutions discussed below are all completely regularized.)  \\
\indent Of course, the notably low sensitivity of the CO map--with the inter-arm regions and the zone surrounding the bar showing little, if any, emission (Figure \ref{fig-5457zoom}; see also \citealt{helfer})--itself raises significant doubts about whether the pattern speed can be accurately measured.  
According to Paper I, however, as it is engaged with the TWR calculation, regularization essentially attributes information within a given element of the TWR kernel to all elements at the same radius in all other slices, thereby compensating for zones which lack clear signatures of patterns.  (Whether or not this is accurate depends primarily on how well the PA and inclination have been determined.)  Errors that may be expected to arise in such low S/N regions are furthermore smoothly redistributed throughout the solution.  \\
\indent Given that the net result is an effective extraction of the available information, we argue that our regularized solution is a reliable indication that the data do not support the measurement of more than a single pattern speed.  In fact, when we impose a transition near the expected molecular bar end, solutions with 0.75$<$$r_t$$<$1.6 kpc either indicate no difference between the inner and outer pattern speeds, or the inner speed is only slightly higher than, and within the errors of, the outer.  \\
\indent As evidence of the incompatibility of a distinct, higher pattern speed with the data, in Figure \ref{fig-5457vslice} we show a comparison between the $<$$v$$>$ (defined in Paper II and references therein) reproduced by the best-fit, constant pattern speed solution and by two mock pattern speeds: one for the bar, $\Omega_b$=115 $\kmskpc$ (chosen in order that the bar ends near CR) and one for the spiral, $\Omega_s$=48 $\kmskpc$ (adopted from the value of the best-fit solution).  There, the best-fit solution yields a significantly better fit to the data than the two speeds for slices inside $\vert y\vert$$\sim$1.0 kpc. \\
\indent This conclusion does not appear to depend on the PA or inclination, unlike in M51 where the radial variation in the best-fit solution varies from PA to PA (Paper II).  For all angle combinations a single, constant pattern speed yields a significantly better fit to the data than two, distinct speeds.  
Nor is the measurement sensitive to spatial variation in the CO-H$_2$ conversion factor.  Based on the metallicity gradient -0.028$\pm$0.01 dex kpc$^{−1}$ measured from oxygen abundances \citep{cedres}, we modeled an increase in $X$ with radius across the CO emitting disk (according to the galaxy-to-galaxy scaling of \citealt{arimoto}). In this case a single, constant pattern speed is not only once again the best-fit to the data, but at $\Omega_p$=53$\pm$6 $\kmskpc$, there is virtually no change to the measured value.\\
\indent Our result, namely, that there is no distinct pattern speed over the length of the molecular bar, therefore seems to be authentic.  Observations of HII regions downstream of the molecular spiral arms \citep{waller} also support our finding that the spiral arms lie inside corotation (although our spiral speed is much higher than the value inferred by \citet{waller} from measurements of the difference in position of the H$_{\alpha}$ and CO emission).  The relative spatial offset between the CO and UV emission in Figure 3 of \citet{waller}, while vague in places, seems sustained throughout the zone of the spiral and diminishes roughly near our CR radius.  Since we find that the bar and spiral have the same pattern speed, this further implies that the bar ends well within its corotation radius.    \\
\indent This scenario has been suggested by \citet{ksw} who, in first reporting on the 25$\degree$ offset between the molecular and stellar bar position angles, noted a resemblance to the hydrodynamical simulations of \citet{cg}.  Such a decoupled central gas concentration is found when the stellar bar is slow 
and drives spirals that develop inside corotation, as we suggest here. \\
\indent A slow bar, ending well inside corotation, also seems to be reconciled with central DM content in M101 implied in the multi-component simulations of \citet{pepitas}, which best reproduce the observed velocity field with a minimum disk.  Where the DM contribution to the gravitational potential is large, \citet{debs} find that interaction between a bar and the DM halo through dynamical friction decelerates the bar (\citealt{weinberg}; \citealt{debs}), such that it grows in length disproportional to the greater increase in $r_{CR}$.  It should be noted, however, that the slowness of the bar implied in this case is a matter of debate (cf. \citealt{sellwood}, \citet{dubinski} and \citealt{wk} who argue that slow bars are an artifact of the resolution of the N-body simulations used to model the interaction).  
\subsection{\label{sec:results342}IC 342}
\begin{figure}
\begin{center}
 \leavevmode
\plotone{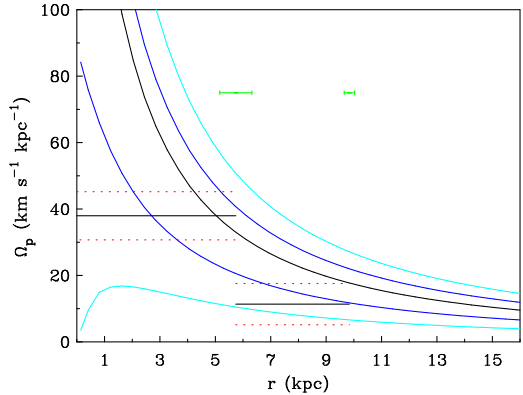}
\end{center}
\figcaption{
Best-fit regularized solution for IC 342 with $r_c$=9.8$\pm$0.19 kpc for PA=42$\degree$$\pm$3$\degree$.  For this solution, bins exterior to $r_c$ (not shown) have been calculated without regularization.  Dashed red lines and green horizontal error bars represent the dispersion in the pattern speeds and in $r_t$ and $r_c$ in solutions derived with a two-pattern speed model at the three PAs.  The values in the zone of the bright spiral structure correspond to $\Omega_{p,1}$=38$\pm$7 $\kmskpc$ out to $r_{t}$=5.7$\pm$0.71 kpc and $\Omega_{p,2}$=11$\pm$6 $\kmskpc$ out to $r_c$.  Curves for $\Omega$, $\Omega$$\pm$$\kappa$/2 and $\Omega$$\pm$$\kappa$/4 are shown in black, cyan and blue.
\label{fig-342twr}}
\end{figure}
\begin{figure}
\begin{center}
\epsscale{1.0}
\plotone{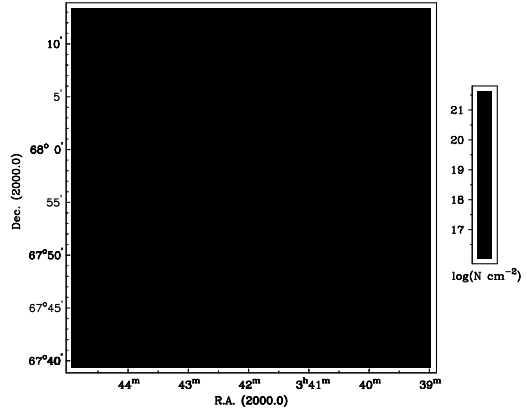}
 \end{center}
    \figcaption{Total H$_2$+HI surface density in IC 342  highlighting the structure inside $\sim$9.8 kpc.  The transition $r_{t}$ marking the extent of the inner pattern speed in our solution is shown as a dashed black circle.  The horizontal bar near the bottom left indicates the physical scale.  
\label{fig-342zoom}}
\end{figure}
The TWR calculation for this galaxy proceeds with the use of a radial bin width $\Delta r$=0.41 kpc (as established by the resolution of the combined maps) using slices out to $\vert y\vert$=13.3 kpc.  The best-fit solution, plotted in Figure \ref{fig-342twr} together with angular rotation and resonance curves, measures two distinct pattern speeds inside the cut radius $r_c$=9.8$\pm$0.19 kpc.  As in M101, we find that this cut radius accurately reflects the location where the distortion/warp in the outer disk begins, e.g. as identified in the HI surface density/velocity field (\citealt{crosthwaiteHI}; see Figure \ref{fig-mom0}).\\
\indent Furthermore, the speed $\Omega_{p,1}$=38$\pm$7 $\kmskpc$ inside $r_t$=5.7$\pm$0.7 kpc is nearly identical to the value \citet{crosthwaiteCO} estimate based on their determination $r_{CR}$=4.2$\pm$0.7 kpc from inspection of the field of velocity residuals.  According to the solution in Figure \ref{fig-342twr}, as parameterized by $r_t$ this pattern terminates at either corotation (here at $r$$\sim$5 kpc; conditional on the rotation curve, which is different from that of \citealt{crosthwaiteCO}) or OLR (within the uncertainties) in favor of the second, lower speed $\Omega_{p,2}$=11$\pm$6 $\kmskpc$.  \\
\indent The transition between the two speeds occurs very near where the spiral structure becomes visibly four-armed (see Figure \ref{fig-342zoom}).  This seems to establish that the outer arms are, in fact, best described by a speed that is distinct from that of the spiral interior, as previously suggested by \citet{crosthwaiteHI}.   
Moreover, Figure \ref{fig-342twr} suggests a possible resonant link between the two speeds in IC 342, in that the lower speed begins near its ILR (or possibly its inner 4:1 resonance).  The low speed also seems to end at the 4:1 resonance, and possibly CR (within the uncertainties).  However, the cut radius $r_c$ in Figure \ref{fig-342twr} seems less an accurate bound on the four-armed spiral, which continues out to $r$$\sim$11 kpc (see the zeroth-moment map in Figure \ref{fig-mom0}), than an indication of where the warp begins.  In this case, whether the warp and four-armed spiral are related or share a common origin is unclear.\\
\indent Although we might also infer from Figure \ref{fig-342twr} that the inner speed lacks an ILR, our uncertainty in the rotation curve at the center is high, given the low resolution of the maps.  Without more confidence in the behavior of the $\Omega$-$\kappa$/2 curve, in particular, it remains unclear whether this speed, and the structure it describes, has an inner bound.  Currently, then, both the spiral structure traced by CO and the distinct bar with length $a_B$$\sim$1.5 kpc \citep{crosthwaiteHI} appear to rotate with speed $\Omega_p$$\sim$40 $\kmskpc$.  As in M101, this notably implies that the bar ends well inside its corotation radius, a circumstance which may also be evidenced by the $\sim$9$\degree$ offset between the molecular and stellar bar major axes \citep{crosthwaiteCO}, as similarly discussed in $\S$\ref{sec:m101inner}.\\
\indent On the other hand, the molecular bar and inner spiral could have different pattern speeds measurable, in principle, but for the width of the radial bins; information from the zone of the molecular gas, which covers a relatively small extent, is limited to only a minor contribution in solutions.  Although the two-pattern speed solution in Figure \ref{fig-342twr} yields a significantly better fit to the data, a higher pattern speed can, in fact, be recognized in solutions modeled with three distinct pattern speeds: the lowest $\chi_{\nu}^2$ solution of this type at PA=42$\degree$ favors a transition at $r_t^i$=1.6$\pm$0.45 kpc from $\Omega_p$=34$\pm$9 $\kmskpc$ to a higher $\Omega_p^i$=70$\pm$12 $\kmskpc$, with $\Omega_p$=12$\pm$4 $\kmskpc$ between $r_t$=5.3$\pm$0.64 kpc and $r_c$=9.8$\pm$0.19 kpc.  (Errors represent the dispersion in three-speed solutions where the inner-most transition, found best at the optimal PA, is held fixed from PA to PA).  This speed is particularly compelling since the bar would end very near CR, and at this radius overlap with the lower, spiral pattern's inner 4:1 resonance. \\ 
\indent However, we emphasize that this identification is inconclusive at this point; the zone of the higher pattern speed is covered by only four radial bins, or less than 11\% of the disk.  (Note that the CO--which has a much higher resolution than the HI--alone cannot be used as a continuity-obeying kinematic tracer, unlike in M101.)  Sensitivity to variation in the CO-to-H$_2$ conversion factor (e.g. given the metallicity gradient in this galaxy measured by \citealt{vila}) is also currently untestable; with such a large radial bin width any variation in $X$ that we might accommodate would be crude at best and, with only four bins covering the CO emission, would likely yield an insignificant result.  
\subsection{\label{sec:results3938}NGC 3938} 
\begin{figure}
\begin{center}
 \leavevmode
\plotone{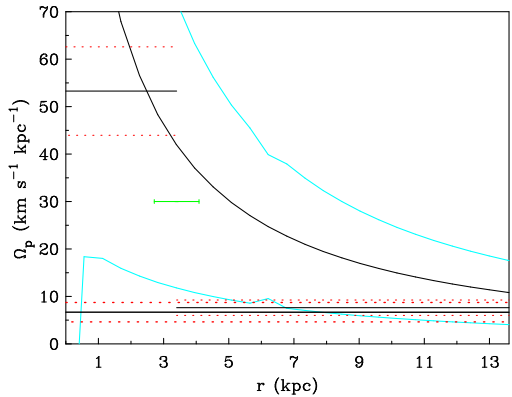}
\end{center}
\figcaption{
Best-fit regularized solutions for NGC 3938 with PA=21$\degree$$\pm$2$\degree$.  Dashed red lines for both solutions, $\Omega_p^A$ (thick line) and $\Omega_b^B$ (thin line) represent the dispersion in the pattern speeds at the three PAs.  The green horizontal error bar shows the dispersion from PA to PA in $r_t$ in solutions derived with a two-pattern speed model.  In the two-speed solution, the value in the inner zone corresponds to $\Omega_{p,1}^B$=53$\pm$9.2 $\kmskpc$ out to $r_{t}$=3.4$\pm$0.7 kpc followed by $\Omega_{p,2}^B$=7.7$\pm$1.6 $\kmskpc$, while the single, constant speed solution measures $\Omega_p^A$=6.6$\pm$2.1 $\kmskpc$.  Curves for $\Omega$ and $\Omega$$\pm$$\kappa$/2 are shown in black and cyan. \label{fig-3938twr}}
\end{figure}
\begin{figure}
\begin{center}
\epsscale{1.0}
\plotone{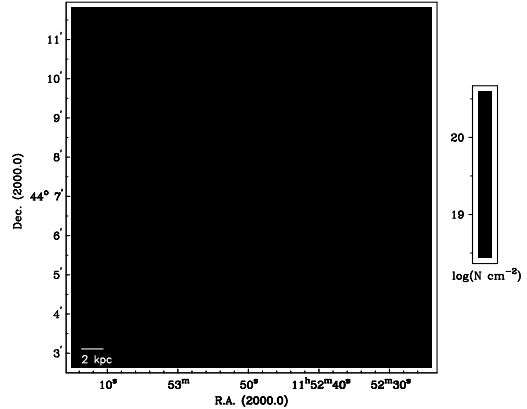}
 \end{center}
    \figcaption{Total H$_2$+HI surface density in NGC 3938 highlighting the structure inside $\sim$15 kpc.  The transition $r_{t}$ marking the extent of the inner pattern speed in solution $\Omega_p^B$ is shown as a dashed black circle.  The horizontal bar near the bottom left indicates the physical scale.  
\label{fig-3938zoom}}
\end{figure}
\indent Given the resolution of the HI data (to which the BIMA CO cube had been smoothed prior to the combination) and, as a consequence, the rather large bin width $\Delta$$r$=1.7 kpc, our TWR solutions calculated in slices $\vert y\vert$$\leq$13.1 kpc contain only 8 radial bins.  This, together with the low inclination of the disk--which compromises the signatures of departures from axisymmetry in the velocity field--limits how well radial variation in the pattern speed can be detected, if present.  Under this scenario, we find that two TWR solutions calculated with regularization over all radial bins (i.e. with no $r_c$) fit the data equally well.  The first ascribes a single, constant pattern speed $\Omega_p^A$=6.6$\pm$2.1 $\kmskpc$ to structure throughout the disk, while the second incorporates the measurement of a distinct, higher pattern speed $\Omega_{p,1}^B$=53$\pm$9.2 $\kmskpc$ inside $r_t$=3.4 kpc (in addition to the lower $\Omega_{p,2}^B$=7.7$\pm$1.6 $\kmskpc$).  \\
\indent The measurement of any pattern speed at all may be surprising, given the low inclination angle and the apparent lack of strong, organized structure in both the HI and (smoothed) CO emission analyzed here.  Comparison with the rotation curves in Figure \ref{fig-3938twr} greatly aids our interpretation of the measurements, although any conclusions that might be drawn are tenuous at best; for this low inclination, rotation curves are highly uncertain (though similar to those throughout the literature, e.g. \citealt{jv} and \citealt{combes}).  \\
\indent With this in mind, we find it noteworthy that the low speed $\Omega_{p}^A$ (or $\Omega_{p,2}^B$) is very near $\Omega$-$\kappa$/2 in the outer disk (Figure \ref{fig-3938twr}; and possibly near $\Omega$-$\kappa$/4, as well). 
This similarity seems reminiscent of a pattern formed by closed, precessing elliptical orbits guided by the near-constancy of $\Omega$-$\kappa/2$. 
But since there is no dominant $m$=2 Fourier component visible in the 21 cm emission or in the NIR (as observed by \citealt{castro}) at these radii, 
it is unclear that this speed is associated with spiral structure, at all.  A lack of spiral streaming motions in the atomic gas (as also observed in the ionized component of the ISM; \citealt{jv}), would furthermore lead us to expect that, for any asymmetry in the surface density, a value much closer to the disk angular rotation frequency $\Omega$ would be measured.  \\
\indent At radii $r$$\lesssim$6 kpc, on the other hand, clear spiral structure is identifiable in the NIR (see the K-band image in \citealt{castro}).  At the very least, this seems to suggest that the two-pattern speed solution $\Omega_p^B$ (with $r_t$ as marked in Figure \ref{fig-3938zoom}) may be more realistic than $\Omega_p^A$; the faster inner speed seems more compatible with structure at these radii than $\Omega_p$$\sim$7 $\kmskpc$ since, even with uncertainty in $\Omega$ at the inner radii and the near-constancy of $\Omega$-$\kappa$/2 at the outer radii, the low pattern speed's ILR would lie at $r$$\gtrsim$5 kpc.  \\
\indent Nonetheless, whether $\Omega_{p,1}^B$ is a reliable measurement of the expected higher pattern speed is difficult to assert.  The transition $r_t$ (see Figure \ref{fig-3938zoom}) lies just inside the edge of the CO emission tracing a two-armed spiral (clear in the unsmoothed BIMA SONG map; \citealt{helfer}), and so the measurement presumably reflects the speed of this pattern.  However, even though this speed appears to cover a quarter of the disk, in general we expect that two radial bins would not supply sufficient leverage on the measurement. (This is arguably the reason why the two TWR solutions are indistinguishable).  Indeed, with so few bins in this zone we are prevented from diagnosing radial variation of order higher than zero.  Furthermore, the accuracy of the measurement, if real, depends on how accurately information about the two speeds can be separated according to the parameterization of the transition $r_t$=3.4 kpc, which here may be severely limited by the size of the radial bin width.  (Note that, as in IC 342, the unsmoothed 6'' resolution CO alone cannot be used as a continuity-obeying kinematic tracer, since ISM is not molecule-dominated over the extent of the CO emission.  Tests for sensitivity to variation in the CO-to-H$_2$ conversion factor are also not well-accommodated with so few, large radial bins). \\
\indent These shortcomings notwithstanding, the distinct measurement inside $r$$\sim$3 kpc seems credible 
if only that it is consistent with values indicated by two independent approaches using different tracers of the spiral structure.  
\citet{korchagin} find with a global modal approach that the dominant $m$=3 and $m$=4 modes in their multi-wavelength observation-based simulations are well-described with a pattern speed near $\sim$55 $\kmskpc$.  \citet{mg} propose a similar pattern speed, $\sim$47 $\kmskpc$ (adopted from \citealt{mg} for $D$=11.3 Mpc), based on the azimuthal age gradient across the spiral arms calculated from optical images.  \\
\indent From Figure \ref{fig-3938twr} we might also argue that, within the uncertainties, $\Omega_{p,1}^B$ ends near corotation at $r_{CR}$=2.5 kpc and so represents a physically realistic scenario, also recently identified in M51 (Paper II).  
However, this speed might just as reasonably end at OLR, depending on the rotation curve, the determination of $r_t$, and the measurement itself.  Also, while this pattern appears to lack an ILR, this is uncertain given that the angular velocities are the least well-determined at the inner radii.  At this point there also does not seem to be a clear relation between the two pattern speeds (e.g. overlapping CR and ILR; \citealt{masset2}), especially since resonances for the outer speed $\Omega_{p,2}^B$ are difficult to establish, given the flatness of the resonance curves.\\
\indent While our confidence in the TWR measurement is tempered by the low inclination of the disk, the value of this analysis lies in the prospect of renewed perspective on the nature of the structure in this galaxy.  Although in no way definitive (based on the TWR method, alone), the lower measurement near $\Omega$-$\kappa$/2 (either $\Omega_{p}^A$ or $\Omega_{p,2}^B$), in particular, may be compatible with   
several pieces of evidence that point to the influence of the DM halo first contemplated by \citet{vs}.  \\  
\indent As investigated by \citet{frenk} or by \citet{jogm1}, for example, the DM halo may be responsible for the appearance of structure in this isolated galaxy, which otherwise seems difficult to reconcile with the finding that the gas disk is everywhere subcritical to gravitational instability (as remarked upon by \citealt{combes}, and 
aside from an alternative increase in instability possible through the coupling between multiple components in the system, e.g. \citealt{jogstab}).  \\
\indent Specifically, where the Toomre stability parameter in the gas is high, \citet{bureau} and \citet{frenk} suggest that the torque due to a triaxial halo (predicted by Cold Dark Matter simulations of hierarchical structure formation; e.g. \citealt{dc}) with slow figure rotation can drive structure in extended HI disks.  Material (though not exclusive) to this scenario, the disk of NGC 3938 exhibits intrinsic ellipticity: as similarly diagnosed in other eccentric nearby spirals by \citet{andersen}, the photometric and kinematic position angles of NGC 3938 are offset by nearly 50$\degree$ \citep{daigle}.  In fact, \citealt{castro} measure an ellipticity $\epsilon_J$=0.11 in the J-band.  So while the ellipticity may have arisen with an asymmetric accretion of matter, for example, it could also reflect an $m$=2 potential perturbation occasioned by triaxiality in the DM halo (see, e.g., \citealt{andersen} and references therein).  In this case, the TWR solution may represent a measure of the rotation speed of such a halo, if the structure in the outer disk (which, extends further in HI than in the optical; \citealt{vs}) arises in the manner considered by \citet{frenk}.  Currently, however, it is not clear that the structure in NGC 3938 is compatible with the strong two-armed patterns found in the simulations of \citet{frenk}.  \\
\indent Alternatively, the overall instability of the disk might be increased by a global mass asymmetry resulting from the $m$=1 perturbation to the halo potential considered by \citeauthor{jogm1} (\citeyear{jogm1}, \citeyear{jogm} and \citeyear{jogv1}), which is also a possible source of the ellipticity.
This latter scenario seems to be compatible with other observable characteristics of this galaxy:
like \citet{tully} and \citet{bournaud}, we find the HI disk to be slightly lop-sided toward the north, and this $m$=1 asymmetry appears in the HI velocity field, as well.  (As previously demonstrated with the H$\alpha$ observations of \citealt{daigle}, we measure a $\sim$10-20 km s$^{-1}$ difference between the approaching and receding sides beyond $r$$\sim$11 kpc.)  In this case, the TWR measurement may reflect the response of the gas (and stars) to this perturbation if the elliptical orbits calculated by \citet{jogm1} precess with frequency $\Omega$-$\kappa$/2.\\
\indent Again, our TWR measurement does not confirm, or distinguish between, these possibilities.  In either of the two scenarios it is also not clear if, or how, the inner, higher speed may relate to the lower speed.  
Higher resolution observations are necessary to first establish which of the two solutions, $\Omega_p^A$ or $\Omega_p^B$, is the most appropriate for this galaxy.  Only then will it be possible to more rigorously explore the relation between the two pattern speeds suggested by the solution $\Omega_p^B$ and perhaps then recognize the true nature of the low speed.  
\subsection{\label{sec:results3344}NGC 3344}
\begin{figure}
\begin{center}
 \leavevmode
\plotone{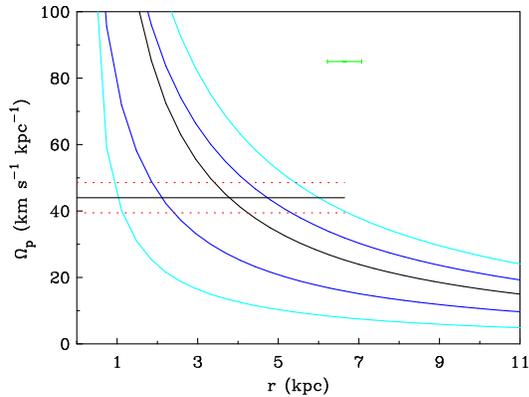}
\end{center}
\figcaption{
Best-fit regularized solution for NGC 3344 with $r_c$=6.8$\pm$0.43 kpc for PA=155$\degree$$\pm$2$\degree$.  For this solution, bins exterior to $r_c$ (not shown) have been calculated without regularization.  Dashed red lines and the green horizontal error bar represent the dispersion in the pattern speeds and $r_c$ in solutions derived at the three PAs.  The value in the zone of the bright spiral structure corresponds to $\Omega_{p,1}$=44$\pm$4 $\kmskpc$ out to $r_c$.  Curves for $\Omega$, $\Omega$$\pm$$\kappa$/2 and $\Omega$$\pm$$\kappa$/4 are shown in black, cyan and blue.  
\label{fig-3344twr}}
\end{figure}
\indent The best-fit solution calculated in slices out to $\vert y\vert$=19.4 kpc with radial bin width $\Delta r$=0.75 kpc (corresponding to the $\sim$20'' resolution of the HI cube) is shown in Figure \ref{fig-3344twr} along with rotation and resonance curves.   
Immediately we notice that, according to our determination of the cut radius $r_c$=6.8$\pm$0.43 kpc--near the location where the warp has been identified to begin ($r$$\sim$7 kpc; \citealt{vm})--this spiral pattern ends at OLR, within the uncertainties.     
The inner bound for the single, constant speed $\Omega_p$=44$\pm$4 $\kmskpc$ in Figure \ref{fig-3344twr}, on the other hand, is less clearly well established here.  The paucity of 21 cm emission inside $r$$\sim$1 kpc not only leads us to suspect that 
the measurement may not be valid all the way to the center, as suggested by our solution in Figure \ref{fig-3344twr} (calculated with bins covering all radii $r$$\geq$0), but it also prevents us from constraining the angular rotation curve there (and the position of the ILR) with confidence.  \\
\indent Even so, according to our estimate $r_{ILR}$$\sim$1.0 kpc, the spiral could reasonably begin at the inner ring, as identified in the optical by \citet{vm}.  In this case, with the single pattern speed implied by our TWR solution, both the inner and outer rings, at $r$=1 kpc and $r$=7 kpc \citep{vm}, are arguably associated with spiral's inner and outer Lindblad resonances.  We note that this depends in no small part on the TWR method; the separation achieved by the cut radius between the strong spiral structure and the $m$=1 distortion grants a measurement for the former constant speed which is significantly different from that implied by the traditional TW method, $\Omega_p$=26$\pm$6 $\kmskpc$ (calculated with the same slices, kinematic parameters, and limits of integration as in the TWR calculation).  \\
\indent The model-independent TWR method, moreover, resolves the ambiguity in the study by \citet{vm}.  There, the uncertainty in the rotation curve precluded the reliable association of the inner and outer rings with resonances, and the estimation of a pattern speed \citep{vm}.  With the identification above we can confirm that the outer pseudoring at $r$$\sim$7 kpc originates with the spiral, as opposed to the bar, which \citet{vm} argue is too small to have an OLR at this radius.  \\
\indent The inner ring, on the other hand, seem more likely related to the bar (which it surrounds; \citealt{vm}), given the proximity of these two features.  That we find the spiral's ILR nearly coincident with the location of the inner ring might then be a manifestation of resonance overlap between the speeds of the bar and spiral.  
If the spiral with lower speed is driven at resonance, for example, our TWR solution may suggest a reasonable speed for the bar, which we are prevented from measuring here.
In particular, if we associate the inner, ILR ring with inner extent of the spiral (inside of which the HI density falls off), such a scenario could be realistically achieved via CR-ILR mode-coupling, where the bar rotates with pattern speed $\Omega_p$$\gtrsim$150 $\kmskpc$ for $r_{CR}$$\sim$1.0 kpc.  The bar, with length $a_B$$\sim$0.7 kpc \citep{vm}, in this case would be reasonably fast, with $r_{CR}/a_B$$\sim$1.4.  Alternatively, we might expect a bar pattern speed $\Omega_p$$\gtrsim$80 $\kmskpc$, 
in the event of the CR-inner 4:1 resonance overlap discussed in $\S$\ref{sec:results5457}, in which case  $r_{CR}/a_B$$\sim$3.5.  In the former (latter) scenario, the inner ring could be located near the bar's CR (inner 4:1 resonance), at least with the rotation and resonance curves derived here.  \\
\indent 
Our TWR solution also seems to endorse the scenario speculated upon by \citet{vm} for the origin of the warp in this galaxy, namely through the non-linear coupling proposed by \citet{masset1}.  In this scheme, the warp (manifest in the lop-sided part of the HI distribution, which extends 34\% further toward the SE than NW; \citealt{vm}) would originate through a coupling at OLR between a spiral density wave and two ``warp waves''.  Since we have established that the spiral structure ends at OLR, in the absence of tidal interaction with a companion galaxy \citep{thilker}, the spiral pattern itself could be capable of generating the warp in the outer disk.
\section{\label{sec:summ}Summary and Conclusions}
Direct pattern speed measurement affords an observational resolution for several fundamental issues in the nature and origin of spirals.  The relation between bar and spiral pattern speeds, the number and radial domains of patterns speeds that can be sustained in the disk, and whether spiral structure is steady or winding, for example, can all be established with knowledge of spiral speeds and their radial variation.  The TWR method, a technique for measuring radially varying pattern speeds, supplies us with the first such measurements to address these issues.\\
\indent In this paper we have applied the TWR method to observations of CO and HI in four spiral galaxies.  For this work, we have expanded the number of spiral galaxies to which TW-type calculations can be applied by considering, for the first time, the combination of both molecular and atomic gas.  
Together, the two ISM phases better meet the continuity requirement of the method and also afford greater insight into whether (and how) multiple spiral pattern speeds extend over a large range of radii.  This has notably increased the sample size of galaxies analyzed with the TWR method so far.\\
\indent Our TWR solutions for M101 ($\S$\ref{sec:results5457}), which, of all the solutions presented here, are equipped with the smallest radial bins, very clearly show radial variation in the pattern speed across the disk.  Within the inner 3', we find convincing evidence that the bar and spiral have the same pattern speed (distinct from an exterior speed), and that both lie inside corotation.  This represents a scenario quite different from recent findings in favor almost exclusively of fast bars (with 1$\lesssim$$r_{CR}/a_b$$\lesssim$1.4; \citealt{corsini}) albeit in mostly early-type barred galaxies.\\
\indent In M101--as well as in IC 342 and possibly NGC 3938--we also find that the extensive spiral structure there is best described with more than a single pattern speed.  Furthermore, in both M101 and IC 342 we find evidence that the transition between two speeds coincides with resonance overlap.  In the former, the transition between the two- and four-armed patterns occurs near the overlap of the inner's OLR and the outer's inner 4:1 resonance.  In the latter case, the inner pattern transitions to a four-armed spiral at an OLR-ILR overlap.  \\
\indent Although the specific resonance overlaps identified in M101 and IC 342 have not been conclusively demonstrated as true instances of mode-coupling (e.g. \citealt{syg}, \citealt{masset2}), together with the CR-inner 4:1 resonance overlap in M51 (Paper II), this work suggests that there exist several possibilities dictated by resonance overlap by which extensive spiral structure can be sustained.  Our findings are qualitatively similar to the barred spiral simulations of \cite{rs}, but we note that the spatially coincident existence of multiple modes with different pattern speeds, as often found there beyond the bar, is untested with the TWR calculation here.  Even so, the outer four-armed spirals in both M101 and IC 342 are arguably excited at resonance by the patterns interior. \\
\indent On the other hand, in our sample we also find extensive spiral structure in the absence of clear resonance overlap.  The speed characteristic of the outer, flocculent structure in NGC 3938, for instance, does not seem to be clearly related to that of the inner, two-armed spiral, if the two speeds are distinct.  
In addition, the tight spiral structure throughout the HI emitting disk in NGC 3344 is best described with a single, constant pattern speed.  \\
\indent In no case do we find that pattern speed is a smoothly decreasing function of radius, as might be expected for a winding spiral.  Our cut procedure, however, may introduce a bias against such spirals if they exist preferentially in the outer regions of the disks in our sample; the radial bins covering the outer portion of three of the four galaxies where a warp is evident are excluded from our models and calculated without regularization.  \\
\indent Even though in this case we cannot characterize the patterns that may distinguish the warped regions of the disks in our sample, accurate measurements for the pattern speeds of the structure interior can, themselves, recommend different mechanisms by which warps are expected to originate and evolve.  
The galaxies in this sample are consistent with various existing explanations.  Where the spiral structure ends at OLR in a warped, isolated galaxy (e.g. NGC 3344), for example, spirals themselves may be capable of exciting the warp through the non-linear coupling proposed by \citet{masset1}.  In the absence of both a companion and this resonance boundary, but where the disk is intrinsically elliptical, as in NGC 3938, the ellipticity, the warp and/or the instabilities in the disk may arise with a DM halo or the asymmetric accretion of matter (discussed in $\S$ \ref{sec:results3938}).  
An obvious companion, on the other hand, is a clear, potential source of tidal perturbation to the disk (e.g. M101 and IC 342).  Although an accompanying warp or distortion may not necessarily relate to the spiral structure at resonance, where the speed of outer structure is near the $\Omega$-$\kappa$/2 curve, for example, this structure may be resonantly excited (at ILR) by the external perturber (see e.g. \citealt{salo}).\\
\indent These issues, and the existence of extensive spiral structure, are best developed with a larger sample of galaxies, where meaningful trends will be more clearly established.  For greatest effect, the TWR calculation should be applied to higher resolution HI observations than we analyze here.  In all of the galaxies in our sample (excepting M101) the resolution of the HI data limits our confidence in how well radial variation is constrained.  It also inhibits our leverage on information extracted in the zone traced by CO emission (which we smooth to the resolution of the HI in combining the two observations). \\
\indent  A large radial bin width likewise prevents us from satisfactorily testing the effect of variation in the CO-to-H$_2$ conversion factor on TWR pattern speed estimates.  Future studies with higher resolution observations (and therefore more radial bins across the CO-emitting disk) may be able to better test for the effects of a gradient or arm-interarm variations in $X$.  We note, though, that these types of variations were found to have a negligible effect on TW estimates \citep{zrm04}, and here we find that low-level variation in the $X$-factor produces very little change in our TWR solutions for M101. \\ 
\indent In general, these types of TWR measurements in nearby galaxies are invaluable for interpreting observational studies of the evolution of bar and disk parameters now possible with HST.  They also promise to be significant for studies at intermediate redshift, and future studies with JWST, which will extend structural parameter measurement to larger redshifts and allow smaller bars and the earlier evolution of disks to be studied.

We would like to thank Pertti Rautiainen for his constructive comments on this manuscript. This material is based on work partially supported by the National Science Foundation under grant AST 03-06958 to R. J. R.  

\end{document}